\documentclass[aps,twocolumn,preprintnumbers,prd,superscriptaddress,nofootinbib,10pt,floatfix]{revtex4-2}

\usepackage{times}
\usepackage{amsfonts}
\usepackage{amsmath}
\usepackage{amssymb}
\usepackage{natbib}
\usepackage{graphicx}
\usepackage{enumitem}
\usepackage{xcolor}
\usepackage[colorlinks=true,linkcolor=magenta,citecolor=blue]{hyperref}
\usepackage[nameinlink]{cleveref}
\usepackage[outdir=./]{epstopdf}
\usepackage[normalem]{ulem}
\usepackage{amsmath}
\usepackage[labelfont=bf]{subfig}
\usepackage{ragged2e}
\usepackage{subcaption}   
\DeclareCaptionJustification{justified}{\justifying}
\usepackage{orcidlink}

\usepackage{newunicodechar}
\newunicodechar{⊙}{\odot}

\begin{document}
\definecolor{orange}{rgb}{0.9,0.45,0}
\def\CovDev{D}
\def\Res{{\mathcal R}}
\def\Gammaflat{\hat \Gamma}
\def\metricflat{\hat \gamma}
\def\Dflat{\hat {\mathcal D}}
\def\part_n{\partial_\perp}
%
\def\Lie{\mathcal{L}}
\def\A{\mathcal{X}}
\def\Aphi{\A_{\phi}}
\def\hAphi{\hat{\A}_{\phi}}
\def\E{\Phi}
\def\Ham{\mathcal{H}}
\def\M{\mathcal{M}}
\def\R{\mathcal{R}}
\def\p{\partial}
\def\hg{\hat{\gamma}}
\def\hA{\hat{A}}
\def\hD{\hat{D}}
\def\hE{\hat{E}}
\def\hR{\hat{R}}
\def\hcA{\hat{\mathcal{A}}}
\def\hDelt{\hat{\triangle}}
\def\na{\nabla}
\def\dif{{\rm{d}}}
\def\non{\nonumber}

%
\renewcommand{\t}{\times}
\long\def\symbolfootnote[#1]#2{\begingroup%
\def\thefootnote{\fnsymbol{footnote}}\footnote[#1]{#2}\endgroup}
\title{Rotating Fermion-Boson Stars in $R$-squared Gravity}

\author{Saeed Fakhry\orcidlink{0000-0002-6349-8489}} 
\email{saeed.fakhry@uv.es}
\affiliation{Departamento de Astronomía y Astrofísica, Universitat de València, Avinguda Vicent 
Andrés Estellés 19, 46100, Burjassot (València), Spain}

\author{Jorge Castelo Mourelle\orcidlink{0000-0002-0074-2608}} 
\email{jorge.castelo@correo.nucleares.unam.mx}
\affiliation{Instituto de Ciencias Nucleares, Universidad Nacional Autónoma de México, Circuito Exterior C.U., A.P. 70-543, México D.F. 04510, México.}

\author{Nicolas Sanchis-Gual} 
\affiliation{Departamento de Astronomía y Astrofísica, Universitat de València, Avinguda Vicent 
Andrés Estellés 19, 46100, Burjassot (València), Spain}

\author{Daniela Doneva} 
\affiliation{Departamento de Astronomía y Astrofísica, Universitat de València, Avinguda Vicent 
Andrés Estellés 19, 46100, Burjassot (València), Spain}

\author{Stoytcho Yazadjiev} 
\affiliation{Department of Theoretical Physics, Sofia University “St. Kliment Ohridski”, Sofia 1164, Bulgaria}
\affiliation{Institute of Mathematics and Informatics, Bulgarian Academy of Sciences, Acad. G. Bonchev St. 8, Sofia 1113, Bulgaria}

\author{Jos\'{e} A. Font\orcidlink{0000-0001-6650-2634}}
\affiliation{Departamento de Astronomía y Astrofísica, Universitat de València, Avinguda Vicent 
Andrés Estellés 19, 46100, Burjassot (València), Spain}
\affiliation{Observatori Astron\`{o}mic, Universitat de Val\`{e}ncia,
C/ Catedr\'{a}tico Jos\'{e} Beltr\'{a}n 2, 46980, Paterna (Val\`{e}ncia), Spain}

\begin{abstract} 
Fermion-boson stars are compact equilibrium configurations composed of ordinary fermionic matter and a bosonic dark component interacting only through gravity. Such systems provide a natural framework for exploring deviations from standard neutron-star models, including the possible accumulation of dark matter inside neutron stars, and may be relevant for compact objects near the low-mass black-hole gap.  We construct static and uniformly rotating fermion-boson stars within the framework of $R$-squared $f(R)$ gravity, characterized by the functional form $f(R)=R+aR^{2}$, where $a$ is a positive parameter governing the effective mass scale from the scalar degree of freedom. The fermionic sector is modeled as a perfect fluid described by a tabulated equation of state at zero temperature, while the bosonic component is represented by a self-interacting complex bosonic field. Our results show that the scalar degree of freedom modifies the spatial distribution of both the bosonic field and the fermionic pressure, enlarges the domain of admissible equilibrium solutions, and increases the maximum supported masses relative to general relativity. Our models remain compatible with current astrophysical and gravitational-wave constraints, suggesting that fermion-boson stars in $R$-squared gravity offer a promising framework to investigate the combined effects of dark bosonic matter, rotation, and strong-field modifications of gravity in compact objects. 
\end{abstract}


\maketitle
\vspace{0.8cm}

\section{Introduction} 

Compact objects are one of the most promising laboratories for testing gravity in the strong-field regime. In particular, neutron stars (NSs) probe densities, pressures, and spacetime curvatures far beyond those accessible in Solar-System experiments, making them ideal environments for investigating possible deviations from General Relativity (GR). Among the broad class of extensions to Einstein gravity, $f(R)$ theories represent one of the most extensively studied and phenomenologically viable modifications \cite{Sotiriou:2008rp,DeFelice:2010aj}. In these theories, the Einstein-Hilbert action is generalized by replacing the Ricci scalar $R$ with a nonlinear function $f(R)$, thereby introducing an additional scalar degree of freedom in the gravitational sector, also known as ``scalaron" in cosmology \cite{Pi:2017gih}.  Such theories have attracted considerable attention in cosmology and astrophysics because they can reproduce a broad range of observational results while remaining compatible with current gravitational constraints.  Any viable $f(R)$ model must simultaneously satisfy weak-field and Solar-System constraints while still allowing potentially observable deviations in the strong-field regime of compact objects.

Several viable classes of modified gravity theories have been proposed as extensions of GR, including a broad family of $f(R)$ models \cite{1980PhLB...91...99S,2004GReGr..36.1765N,2007PhRvD..76f4004H,2008CQGra..25m5002A,Sotiriou:2008rp,DeFelice:2010aj}. While some of these models require screening mechanisms to satisfy Solar-System and cosmological constraints, they remain highly valuable in strong-field astrophysics as a representative class of modified theories of gravity with a finite range of the additional (scalar) force \cite{2018PhRvD..97f4016S}. Thus, they allow for a systematic testing of gravity via compact objects.

In an astrophysical context, compact stars in $f(R)$ gravity have been used to understand how the scalar degree of freedom modifies the structure and observable properties of relativistic stars \cite{Yazadjiev:2014cza, 2014JCAP...10..006S, Yazadjiev:2015zia,2022PhLB..82636929N,2023PhRvD.107j4019N}. The majority of studies in this area have focused on static or slowly rotating perfect fluid configurations, which have been employed to study deviations in the mass-radius relation, stellar compactness, internal matter distribution, moments of inertia, and maximum supported masses. Rotation is an essential ingredient of realistic NS modeling. The reason is that all observed NSs rotate, and the fastest known pulsars reach spin periods of $\sim 1.4\,\mathrm{ms}$, corresponding to dimensionless spins $\chi \sim 0.4$~\cite{2006Sci...311.1901H}. Even more rapidly rotating NSs, e.g. close to the mass-shedding limit, can naturally arise from binary mergers or core-collapse events. On the other hand, rotating stars provide a stringent framework for testing modified gravity, since rotation introduces frame dragging, rotational deformation, and an increase of the maximum supported mass toward the mass-shedding limit, all of which can interact non-trivially with the scalar degree of freedom in the strong-field regime. Slow rotation approximation, in linear order with respect to the rotational frequency, is reliable for $\chi \lesssim 0.1$ and becomes increasingly inaccurate for the rapid rotators relevant to astrophysics~\cite{2024PhRvD.110h4027A}. This motivates the construction of fully relativistic uniformly rotating equilibrium models~\cite{Yazadjiev:2015zia,Doneva:2015hsa}, rather than extrapolations from static or slowly rotating configurations.

Most of the compact-star studies cited above in fact concern the quadratic model $f(R)=R+aR^2$, commonly referred to as $R$-squared gravity, where $a$ is a constant parameter controlling the deviation from GR. In the present work, we adopt this $f(R)$ model. It is dynamically equivalent to a scalar-tensor theory with a massive scalar field \cite{2006CQGra..23.5117S}, making it a useful framework for probing strong-field deviations from GR with a final range scalar force. Within this model, additional studies have examined static and slowly rotating configurations in the Palatini formalism \cite{2017GReGr..49...25T}, static boson stars~\cite{maso2021boson,maso2023birth,maso2024numerical}, anisotropic fluid models \cite{2018PhRvD..97l4009F}, inequivalent definitions of gravitational mass \cite{2020PDU....2700411S}, extensions with matter-curvature couplings \cite{2021JCAP...08..055P}, and uniformly rotating stars with axion couplings \cite{2020MNRAS.498.3616A}, as well as related strong-field phenomenology in scalar-tensor gravity \cite{2018EPJC...78..586S,Reyes:2025oet,2026arXiv260516506O}. These works indicate that $R$-squared modifications can produce measurable changes in the mass-radius relation, compactness, and global observables such as the moment of inertia, motivating a further study of compact star models within this modified gravity model.

The observation of unusually massive compact objects, such as the secondary component of GW190814 and pulsars with masses exceeding~$2 M_{\odot}$ \cite{DiGiovanni:2021ejn}, has motivated extensive studies of the high-density equation of state (EOS) of NS matter. While sufficiently stiff nuclear equations of state can support such masses, constraints from gravitational-wave observations tend to favor relatively soft behavior at intermediate densities, suggesting a possible tension between different observational requirements. The potential presence of compact objects within the black-hole low-mass gap \cite{LIGOScientific:2024elc} further blurs the distinction between the most massive NSs and the lightest black holes \cite{Ye:2024wqj, Gupta:2019nwj}. These considerations motivate alternative compact-star scenarios, including dark-matter admixed NSs, where an additional bosonic component can modify the mass–radius relation and increase the maximum supported mass without requiring an excessively stiff nuclear EOS.

In this regard, the possibility that compact stars may contain an additional dark component has attracted considerable attention in recent years. Such scenarios are motivated by the expectation that NSs can accumulate dark matter during their evolution, potentially leading to observable modifications of their structure and global properties. Static realizations of such systems have been studied in several scenarios within GR, including two-fluid models, mirror dark matter, and configurations in which the dark component is described by bosonic scalar or vector fields ~\cite{HENRIQUES198999,Henriques:1989ez,Henriques:1990xg, 2011PhRvD..84j7301L, 2020arXiv201101318D, Kain:2021bwd, 2021PhRvD.103d3009K, Jockel:2023rrm,  2026PhRvD.113f3014G}. In parallel, a number of works have explored through numerical relativity the dynamical evolution and formation channels of these systems, investigating whether the dark component can remain gravitationally bound to the fermionic sector and form long-lived compact configurations \cite{PhysRevD.87.084040, DiGiovanni:2020frc,DiGiovanni:2021vlu,Karkevandi:2021ygv, Giangrandi:2022wht, Nyhan:2022pda, Alvarez-Rios:2024kfr,Lazarte:2025etl}. Such analyses are essential for assessing whether mixed configurations can arise naturally and persist as astrophysically relevant compact objects, see, e.g, \cite{2023ApJ...953..115G, 2022EPJWC.27407009S, 2025arXiv250420825G, 2025A&A...697A.220T}. 

More recently, rotating configurations of mixed compact objects have also begun to be studied within GR. In this rapidly developing field, different descriptions of the dark sector, distinct rotational proposals, and a variety of EOS for the fermionic sector and bosonic-field models have been considered \cite{DiGiovanni:2022mkn, Mourelle:2024qgo,Kumar:2026ddm,Konstantinou:2024ynd,Cipriani:2025tga,Shawqi:2025cca, 2025arXiv251205898Cf}. Rotation can substantially modify the mass, radius, compactness, and other global properties of relativistic stars, while simultaneously introducing new families of equilibrium solutions that are absent in the static limit. In parallel, the stability and oscillation properties of mixed stars have also attracted increasing attention, see, e.g., \cite{PhysRevD.87.084040,2023PDU....4001185A, 2024JCAP...12..042T, 2025MNRAS.544.3549S, 2025PhRvD.111l3013S, 2025arXiv251201641Z, 2026arXiv260503305K}. These effects may have important implications for electromagnetic and gravitational-wave observations. Within this broader context, it is therefore natural to investigate how modified gravity affects the structure and properties of rotating mixed compact objects. In particular, the interplay between rotation, bosonic matter, and scalar field dynamics may lead to observable signatures that differ significantly from those predicted in GR.

Earlier studies have explored different parts of this problem separately. Within GR, static and rotating fermion-boson stars have been constructed with different bosonic potentials, rotation prescriptions, and nuclear EOSs~\cite{DiGiovanni:2021ejn, Mourelle:2024qgo, 2022PhRvD.105f3005D}. In modified gravity, static compact stars with an admixed dark component have been studied in $f(R)$ gravity \cite{2018PhRvD..97b4030L}, dilatonic gravity \cite{2000math......4108B}, Horndeski theory \cite{2022CQGra..39d4001R}, and other extensions \cite{2024EPJC...84..994R, 2026Tangphati}. However, a self-consistent treatment combining modified gravity, rotation, and a mixed fermion-boson sector has not yet been presented. Because each of these ingredients can independently modify compact-star phenomenology, their combined influence may lead to qualitatively new equilibrium configurations and potentially observable strong-field signatures.

In this work, we construct static and uniformly rotating equilibrium solutions for mixed compact stars composed of a standard fermionic NS component and a bosonic sector described by a complex field, considering both configurations in GR and within $R$-squared gravity. The structure of this paper is as follows: In \Cref{sec:ii} we present the theoretical framework, including the formulation of $R$-squared gravity in the Einstein frame and the description of the fermionic and bosonic matter sectors. 
\Cref{sec:iii} presents the approach followed to build rotating fermion-boson stars, including numerical aspects. \Cref{sec:iv} discusses global physical quantities of rotating mixed stars in $R$-squared gravity. In \Cref{sec:v} we present the numerical results for static and rotating models, with emphasis on the impact of the scalar degree of freedom on the equilibrium structure and on the mass-radius relations. Finally, in \Cref{sec:vi} we summarize our conclusions and discuss future perspectives. The paper also contains four appendices. Those include explicit information on the set of field equations to solve (both in the static limit and for rotating configurations) and present additional stellar models we have built taht are not discussed in the main text.  Throughout this work we use geometrized units, $G=c=1$.

\section{FORMALISM}\label{sec:ii}

\subsection{$R$-Squared Gravity}
The action of $f(R)$ gravity is given by
\begin{equation}
S = \frac{1}{16\pi}\int d^4x \sqrt{-g}\, f(R)
+ S_{\rm matter}(g_{\mu\nu},\chi_{\rm matter}),
\label{gravity_theroy_action}
\end{equation}
where $f(R)$ is an arbitrary function of the Ricci scalar $R$ constructed from the spacetime metric $g_{\mu\nu}$. $S_{\rm matter}$ denotes the matter action, and $\chi_{\rm matter}$ collectively represents the matter fields. In order to avoid ghost degrees of freedom and tachyonic instabilities, viable $f(R)$ models must satisfy the conditions ${\rm d}f/{\rm d}R>0$ and ${\rm d}^2f/{\rm d}R^2\geq0$. In the present work, we consider the so-called $R$-squared model of gravity, i.e.,
\begin{equation}
f(R)=R+aR^2,
\label{f_of_r}
\end{equation}
where $a$ is a non-negative constant parameter controlling the deviation from GR.

It is well established that $f(R)$ gravity admits an equivalent scalar-tensor representation \cite{2006CQGra..23.5117S}, corresponding to a Brans-Dicke model with $\omega_{\rm BD}=0$, whose dynamics follow from the action

\begin{eqnarray}
S = \frac{1}{16\pi} \int d^4x \sqrt{-g} [\Phi_{f(R)} R - U(\Phi_{f(R)})] \nonumber\\
+S_{\rm matter}(g_{\mu \nu}, \chi),
\end{eqnarray}
where the gravitational scalar field $\Phi_{f(R)}$ and its potential $U(\Phi_{f(R)})$ are defined as 
\begin{equation}
\Phi_{f(R)} = \frac{{\rm d}f(R)}{{\rm d}R},
\end{equation}
and
\begin{equation}
U(\Phi_{f(R)}) = R \frac{{\rm d}f}{{\rm d}R} - f(R),
\end{equation}
respectively. For the $R$-squared gravity model, these expressions reduce to
\begin{equation}
\Phi_{f(R)} = 1 + 2 a R,
\end{equation}
with the scalar-field potential given by
\begin{equation}
U(\Phi_{f(R)}) = \frac{1}{4a}(\Phi_{f(R)} - 1)^2 .
\end{equation}
It is often convenient to analyze scalar-tensor theories in the Einstein frame, defined via the conformal transformation
\begin{equation}
{g}^{\ast}_{\mu\nu} = \Phi_{f(R)}\, g_{\mu\nu}.
\end{equation}
In this frame, the action takes the form
\begin{eqnarray}
S = \frac{1}{16\pi} \int d^4x \sqrt{-{g}^{\ast}} \left[{R}^{\ast} - 2{g}^{\ast\mu\nu}\partial_\mu\varphi\partial_\nu\varphi - V(\varphi)\right] \nonumber \\
+ S_{\text{matter}}(e^{-\frac{2}{\sqrt{3}}\varphi}{g}^{\ast}_{\mu\nu}, \chi), \hspace{0.5cm}
\end{eqnarray}
where ${R}^{\ast}$ denotes the Ricci scalar constructed from the Einstein-frame metric ${g}^{\ast}_{\mu\nu}$, while the canonically normalized scalar degree of freedom $\varphi$ is introduced through
\begin{equation}
\varphi = \frac{\sqrt{3}}{2}\ln \Phi_{f(R)} .
\end{equation}
The Einstein-frame potential is then expressed as
\begin{equation}
V(\varphi) = A^{4}(\varphi)\, U\bigl(\Phi_{f(R)}(\varphi)\bigr),
\end{equation}
where the conformal factor satisfies
\begin{equation}
A^{2}(\varphi) = \Phi^{-1}_{f(R)}(\varphi) = e^{-\frac{2}{\sqrt{3}}\varphi} .
\end{equation}
For the $R$-squared gravity, this leads to the explicit form
\begin{equation}
V(\varphi) = \frac{1}{4a}\left(1 - e^{-\frac{2}{\sqrt{3}}\varphi}\right)^{2}.
\end{equation}

By varying the action with respect to the metric ${g}^{\ast}_{\mu\nu}$ and the scalar degree of freedom $\varphi$, one can obtain the corresponding field equations in the Einstein frame:
\begin{eqnarray}
{G}^{\ast}_{\mu\nu}=8\pi {T}^{\ast}_{\mu\nu}+2\partial_{\mu}\varphi\partial_{\nu}\varphi &-&{g}^{\ast}_{\mu\nu}{g}^{\ast\alpha\beta}\partial_{\alpha}\varphi\partial_{\beta}\varphi\nonumber\\
&-&\frac{1}{2}V(\varphi){g}^{\ast}_{\mu\nu},
\label{Einstein:tens}
\end{eqnarray}

\begin{eqnarray}
{\nabla}^{\ast}_{\mu} {\nabla}^{\ast\mu} \varphi-\frac{1}{4}\frac{dV(\varphi)}{d\varphi}=-4\pi \kappa(\varphi){T}^{\ast},
\label{Klein-Gordon}
\end{eqnarray}
where in $R$-squared gravity
\begin{equation}
\kappa(\varphi)=\frac{d \ln A(\varphi)}{d \varphi} = -\frac{1}{\sqrt{3}}.
\end{equation}
Here, ${T}^{\ast}_{\mu\nu}$ denotes the total energy-momentum tensor in the Einstein frame, related to its Jordan-frame counterpart through ${T}^{\ast}_{\mu\nu}=A^{2}(\varphi)\,T_{\mu\nu}$.   
Also, ${T}^{\ast}=A^4(\varphi) T$ represents the trace of the corresponding energy-momentum tensor. 

\subsection{Matter content}

In addition to the scalar degree of freedom contribution arising from the underlying $R$-squared gravity, our system also includes a mixed matter configuration composed of a fermionic fluid and a bosonic sector described by a complex field. Both components are minimally coupled to gravity and are assumed to interact exclusively through the spacetime geometry, with no direct nongravitational interaction between them.

Accordingly, the matter energy-momentum tensor contains two independent contributions, one associated with the fermionic fluid and the other with the bosonic field:
\begin{equation}
{T}^{\ast}_{\mu\nu}={T}^{\ast f}_{\mu\nu}+{T}^{\ast b}_{\mu\nu}.
\end{equation}

The NS component is described using the standard relativistic perfect-fluid approximation, as is customary in NS modelling, with energy-momentum tensor given by
\begin{equation}
{T}^{\ast f}_{\mu\nu}=({p}^{\ast}_f+{\rho}^{\ast}_f){u}^{\ast \mu}{u}^{\ast \nu}-{p}^{\ast}_f{g}^{\ast \mu\nu},
\label{SE_f}
\end{equation}
where ${p}^{\ast}_f$ and ${\rho}^{\ast}_f$ are the fermionic pressure and energy density in the Einstein frame, respectively, which can be mapped to the Jordan frame as ${p}^{\ast}_f=A^{4}(\varphi)p_f$ and ${\rho}^{\ast}_f=A^{4}(\varphi)\rho_{f}$.\footnote{Note that the total energy density is denoted by $\rho_f = \rho_0 + \rho_i$, where $\rho_0$ is the rest-energy density and $\rho_i$ represents the internal energy density.}
The fluid 4-velocity $u^{\mu}$ is given by
\begin{equation}
u^{\mu}=\left(u^t,0,0,\Omega u^t \right),
\label{fouthvelocity}
\end{equation}
which can be related to the corresponding one in Einstein frame as ${u}^{\ast}_{\mu}=A^{-1}(\varphi)u_{\mu}$. We also define the zero angular momentum observer velocity as $\upsilon=(\Omega - \sigma)r \sin \theta e^{-\rho}$.
The thermodynamic closure of the baryonic sector is provided by an EOS relating the pressure and the density. In this work we use the AkmalPR EOS in tabulated form~\cite{1998PhRvC..58.1804A}. 

The bosonic component is described by a complex field and represents the matter sector usually associated with a boson star configuration  \cite{PhysRev.172.1331,PhysRev.187.1767,PhysRev.148.1269,Liebling:2012fv,Lai:2004fw,Schunck:2003kk}. This field is minimally coupled to gravity and contributes to the spacetime geometry through its energy-momentum tensor. The action for the Einstein-Klein-Gordon system in the Einstein frame can be written as \cite{2025PhRvD.112l4086H}
\begin{eqnarray}
S_{\rm B} &=& -\frac{1}{2}\int d^4x\sqrt{-{g}^{\ast}}\left[A^2(\varphi){g}^{\ast \mu\nu}{\nabla}^{\ast}_\mu\Phi^{(\star)}{\nabla}^{\ast}_\nu\Phi + \right. \notag \\
&&  \hspace{2.8cm} \left. + A^4(\varphi) V_{\rm s}(|\Phi|^2)\right],
\label{action_bs}
\end{eqnarray}
where $\Phi$ is the complex field and $\Phi^{(\star)}$ denotes its complex conjugate. Depending on the space-time framework, the ansatz for the field should vary.  The potential $V_{\rm s}$ is assumed to depend only on the modulus of the bosonic field, thereby preserving the global $U(1)$ symmetry of the model. In the present work, we adopt a self-interaction potential for the bosonic field of the form
\begin{equation}
V_{\rm s}(|\Phi|^2)=\mu_{b}^2|\Phi|^2+\frac{1}{2}\lambda|\Phi|^4,
\end{equation}
where $\mu_{b}$ is the mass of bosonic field and $\lambda$ is the self-interaction coupling constant. Note that this potential is appropriate for the approach followed here (cf.~the RPV approach \cite{Ryan:1996nk,Vaglio:2022flq,Mourelle:2024qgo} which assumes a large self-interaction coupling constant $\lambda$; see Section~\Cref{section_III_b}).

The canonical energy-momentum tensor associated with the complex bosonic field can be obtained from Eq.~(\ref{action_bs}), 
\begin{align}
{T}^{\ast b}_{\alpha\beta}
&=
A^{2}(\varphi){\nabla}^{\ast}_{(\alpha}\Phi^{(\star)}
{\nabla}^{\ast}_{\beta)}\Phi
\nonumber\\
&\quad
-\frac{1}{2}{g}^{\ast}_{\alpha\beta}
\Bigl[
A^{2}(\varphi){\nabla}^{\ast \mu}\Phi^{(\star)}
{\nabla}^{\ast}_{\mu}\Phi
+
A^{4}(\varphi)V_s\left(|\Phi|^2\right)
\Bigr].
\label{SE_b}
\end{align}
The appropriate ansatz for the bosonic field is fixed by the spacetime symmetries. In particular, non-rotating configurations are usually described by a purely harmonic time dependence, whereas rotating axisymmetric configurations also require an azimuthal dependence in order to be able to obtain stationary regular solutions. This choice directly affects the explicit derivatives of the bosonic field and, therefore, the final form of the corresponding effective energy-momentum tensor.

The general expression above remains valid provided that the scalar-field ansatz is chosen consistently with the symmetries of the underlying spacetime geometry. These aspects will be discussed in detail in \Cref{section_III_b}, together with the regime of large self-interaction coupling constant $\lambda$. In this limit, the energy-momentum tensor of the bosonic field can be approximated by that of an effective perfect fluid, allowing the bosonic sector to be described in terms of effective pressure $p_b$ and energy density $\rho_b$. Although this approximation is particularly useful in the rotating case, we also examine its validity in the static limit. We also note that in analogy to the fermionic sector, the effective bosonic pressure and energy density in the Jordan and Einstein frames are related through the conformal transformations.

Finally, varying the action in Eq.~(\ref{action_bs}) with respect to the bosonic field leads to the Klein-Gordon equation governing the dynamics of the bosonic field. In covariant form, within the Einstein-frame formulation of $R$-squared gravity, the modified Klein-Gordon equation takes the form
\begin{equation}
{\nabla}^{\ast}_{\mu}\left(A^2(\varphi){\nabla}^{\ast \mu}\Phi\right)=A^4(\varphi)\frac{dV_s}{d|\Phi|^2}\Phi.
\end{equation}

\section{Framework}\label{sec:iii}

We now discuss the specific theoretical and numerical approaches we follow to build rotating fermion-boson stars. The static limit is reported separately in Appendix~\ref{app:static_spacetime}. Moreover, the explicit field equations for the rotating case are given in Appendix~\ref{appx_source}.

\subsection{Rotating spacetime}
\label{section_III_b}
The general form of the metric in spherical coordinates $(r,\theta,\psi$) for stationally axisymmetric fermion-boson configurations  can be written in the following way
\begin{equation}
\begin{split}
ds^2 ={}& -e^{\gamma+\rho}dt^2
+ e^{\gamma-\rho}r^2\sin^2\theta
\left(d\psi-\sigma dt\right)^2  \\
&+ e^{2\alpha}\left(dr^2+r^2d\theta^2\right),
\end{split}
\label{metric1}
\end{equation}
where the metric potentials \(\gamma\), \(\rho\), \(\alpha\), and \(\sigma\)
depend on the coordinates \(r\) and \(\theta\).

Regarding the fermionic sector, the fluid 4-velocity can be expressed as a linear
combination of the timelike and azimuthal Killing vectors, denoted by
\(\xi_{(t)}^{\alpha}\) and \(\xi_{(\psi)}^{\alpha}\), respectively,
\begin{equation}
u^{\alpha}
=
u^t\left(\xi_{(t)}^{\alpha}+\Omega\xi_{(\psi)}^{\alpha}\right).
\label{velocity}
\end{equation}
The normalization condition \(u^\alpha u_\alpha=-1\) determines the temporal component of the four-velocity as
\begin{equation}
u^t
=
\left[
-g_{\alpha\beta}
\left(\xi_{(t)}^{\alpha}+\Omega\xi_{(\psi)}^{\alpha}\right)
\left(\xi_{(t)}^{\beta}+\Omega\xi_{(\psi)}^{\beta}\right)
\right]^{-1/2}.
\end{equation}

In coordinates adapted to the Killing vectors \(\xi_{(t)}^{\alpha}\) and
\(\xi_{(\psi)}^{\alpha}\), the angular velocity of the fluid, as measured by an observer
at rest at spatial infinity, is given by
\begin{equation}
\Omega
\equiv
\frac{u^{\psi}}{u^t}.
\end{equation}

It is worth noting how the relevant matter quantities transform between the 
Jordan and Einstein frames under the conformal rescaling 
$g^*_{\mu\nu} = A^{-2}(\varphi)\,g_{\mu\nu}$.
The fermionic energy density and pressure transform as
$\rho_f^* = A^4(\varphi)\,\rho_f$ and $p_f^* = A^4(\varphi)\,p_f$,
whereas the angular velocity $\Omega$ and the coordinate fluid velocity 
$\upsilon$ are conformally invariant and therefore take the same value in both 
frames \cite{Doneva:2013qva}.
This is a consequence of $\Omega$ and $\upsilon$ being ratios of metric 
components or velocity components that scale identically under the conformal 
transformation, so that no additional factor of $A(\varphi)$ appears in their 
definitions. The same invariance holds for the angular velocity of the bosonic 
field $w$ and the frame-dragging potential $\sigma$. 
In the equations that follow, all matter variables are understood as 
Einstein-frame quantities unless stated otherwise.
For the bosonic sector, the field has a two-dimensional distribution depending on both radial and angular coordinates, together with a harmonic dependence consistent with rotating solutions carrying nonzero angular momentum \cite{Herdeiro:2015gia,Ryan:1996nk}, 
\begin{equation}
     \Phi(t,r,\theta,\psi)=\phi(r,\theta)e^{i(\eta\psi-w t)}.
     \label{scalar_field_spinning}
 \end{equation}
 Here, $w \in \mathbb{R}$ is the angular frequency of the field and  $\eta \in \mathbb{Z}$ is the harmonic index (or winding number).

Following~\cite{Mourelle:2024qgo}, our theoretical framework relies on the RVP approach~\cite{Ryan:1996nk,Vaglio:2022flq}, a scheme commonly referred to as the high-coupling-constant approximation. In this approach, one can consistently define effective bosonic energy density and pressure from the bosonic field contribution, so that the field equations describing fermion-boson stars preserve the same structural form as in the purely fermionic case, with additional source terms arising from the bosonic sector. Within the RPV treatment the bosonic component is therefore handled as an effective matter source, with compact support after the practical truncation of the bosonic field tail.
This prescription is physically motivated by the scale separation that appears in the strong self-interaction regime: the bosonic configuration consists of an inner region, where the field amplitude is large and contains almost all the mass, and an outer region where the field decays exponentially. Since the contribution of this tail to the energy-momentum tensor is exponentially suppressed, the RPV scheme neglects it and sets the bosonic source to zero in the outer region, retaining only the compact inner bulge that contains essentially all of the bosonic mass.  This leads naturally to an effective finite-size matter distribution, rather than introducing an independent cutoff by hand. The reader is referred to \cite{Ryan:1996nk,Vaglio:2022flq} for a detailed explanation. Thus, no additional boundary conditions for
the bosonic field are imposed at this stage. The regularity of the metric
potentials and their asymptotic flatness are instead ensured through the
boundary conditions implemented in the Komatsu-Eriguchi-Hachisu (KEH) method \cite{komatsu1989rapidly,friedman2013rotating}, as detailed in the following
section. 

In the RPV framework, although the bosonic field is described by the ansatz in Eq.~\eqref{scalar_field_spinning}, the extended low-amplitude ``tail'' of the field is neglected. More precisely, we assume that the scalar field vanishes in the \textit{outer region}. This condition yields either zero or the value of the scalar field at the bulge, and thus constitutes a central assumption of the approach, namely $|\Phi| = 0$ in the outer region. In contrast, within the inner region, the scalar field is determined by  
\begin{equation}
    |\Phi|^2={\rm{max}}\left[0, \frac{1}{\lambda}\left(\frac{(w-\eta\sigma)^2}{e^{\gamma+\rho}}-\frac{e^{\rho-\gamma}\eta^2}{r^2\sin^2\theta}-\mu_b^2\right)\right].
    \label{ryan_field}
\end{equation}
As a result, the bosonic energy-momentum tensor takes the form of a perfect fluid in the stellar interior, 
\begin{equation}
    T_{\mu\nu}^{b}=(\rho_{b}+p_{b})\hat{u}_{\mu}\hat{u}_{\nu}+p_{b}g_{\mu\nu},
\end{equation}
and is set to zero outside a given radius. The bosonic pressure and energy density read
\begin{equation}
    p_{b}=\frac{1}{4}\lambda |\Phi|^4,\hspace{0.5cm}\rho_{b}=\mu_b^2|\Phi|^2+\frac{3}{4}\lambda |\Phi|^4.
\end{equation}
In addition, the corresponding four-velocity is given by
 \begin{equation}
     (\hat{u}_t,\hat{u}_r,\hat{u}_{\theta},\hat{u}_{\psi})=\frac{(-w,0,0,\eta)}{(\lambda |\Phi|^2+\mu_b^2)^{\frac{1}{2}}} .
 \end{equation}

It should be noted that these quantities will be mapped to their corresponding expressions in the Einstein frame, as done in the previous section, with an appropriate power of the conformal factor $A(\varphi)$.
Moreover, the proper velocity measured by a zero-angular-momentum observer can be specified as
\begin{equation}
    \Bar{\upsilon}=\frac{\eta}{(w-\eta\sigma)}\frac{e^{\rho}}{r\sin\theta} .
\end{equation}
Note that, in general, the fermionic and bosonic components of the star are not assumed to share the same rotation law. The fermionic sector is described as a rigidly rotating fluid, while the bosonic sector generically exhibits an effective differential rotation determined by the field parameters, in particular by $w_s$ and $\eta$ \cite{Ryan:1996nk,Vaglio:2022flq,Adam:2022nlq}. Within the RPV approximation, the truncation of the bosonic field tail makes this effective rotation profile closer to a rigid one, but it does not enforce exact co-rotation with the fermionic component. Since both sectors are coupled only gravitationally, we do not impose any synchronization condition when constructing the initial models. In this context, although the configurations considered here allow the fermionic and bosonic components to rotate at different rates, particular cases in which the rigid angular velocity of the fermionic fluid matches a characteristic angular velocity of the bosonic rotation profile may be of special interest from the perspective of stability. Assessing the stability of these configurations will require dedicated numerical-relativity simulations and is left for future work. We note in passing that~\cite{Lazarte:2025etl} identified a dynamical gravitational synchronization mechanism in static non-rotating fermion-boson stars, where the two components develop a synchronized radial oscillations through gravitational coupling alone. It is possible that analogous synchronization effects may arise in rotating systems, potentially giving rise to interesting observational signatures.

Let us close this section with a summary of the main advantages and limitations of the RPV approximation. On the one hand, the approximation restricts the bosonic sector to a specific class of models, namely a quartic self-interaction potential in the strong-coupling regime. In addition, the exponentially decaying tail of the bosonic field is neglected so that the description focuses on the compact bosonic bulge containing essentially all of the matter content, while the small residual contribution outside this region is truncated. On the other hand, these assumptions bring several advantages. The strong self-interaction regime, together with the truncation of the bosonic field tail, allows the bosonic contribution to be recast effectively in terms of pressure and density variables, making the system structurally closer to a perfect-fluid description. This leads to a substantial numerical simplification of the problem. Moreover, within this approximation the effective differential rotation profile of the bosonic component becomes closer to rigid rotation, improving its compatibility with the fermionic sector considered here. Finally, studies of rotating boson stars indicate that, in the strong quartic self-interaction regime, some non-axisymmetric instabilities may be weakened or even suppressed, suggesting that this regime is particularly relevant for constructing configurations with improved stability properties \cite{DiGiovanni:2020ror,Dmitriev:2021utv,Siemonsen:2020hcg}.

\subsection{Numerical Implementation}
We follow \cite{Vaglio:2022flq} to eliminate the dependence of the bosonic field equations on $\mu_{b}$ and $\lambda$. In geometrized units, where $[\lambda]=[\mathrm{mass}]^{-2}$ and $[\mu_{b}]=[\mathrm{mass}]^{-1}$, $Z_{b}\,\equiv\lambda^{-1/2} \mu_{b}^2$ carries the dimension of mass. This allows us to introduce the following dimensionless variables
\begin{align}
\bar{t} &= Z_{b}^{-1}\,t, \qquad \bar{r} = Z_{b}^{-1}\,r,\\
\bar{p}_{b} &= Z_{b}^2\,p_{b}, \qquad \bar{\rho}_{b} = Z_{b}^2\,\rho_{b}, \qquad \bar{\sigma}=Z_{b}\,\sigma,
\end{align}
where $Z_{b}$ denotes the characteristic bosonic mass scale. It is also convenient to define the dimensionless frequency $\bar{w} = w \mu_{b}^{-1}$. Also, the winding number is rescaled as $\bar{\eta} = \mu_{b}\lambda^{-1/2} \eta$.
In the numerical implementation, the equilibrium equations are solved in terms of the dimensionless variables above. Since $\bar{r}$ is measured in units of $Z_b$, the mass and angular-momentum integrals constructed from $\bar{r}_e$ and the rescaled matter variables return dimensionless quantities $\bar{M}_T=M_T/Z_b$ and $\bar{J}_T=J_T/Z_b^2$. In our case the absolute scale of the fermionic component is fixed by the AkmalPR EOS, while $\mu_b$ and $\lambda$ enter only through the dimensionless bosonic parameters $\bar{w}$, $\bar{\eta}$, and the mass scale $Z_b$. The masses, radii, and angular momenta quoted in \Cref{sec:v} below are therefore given in physical units, not in units of $Z_b$. To fix the absolute scale, we choose the dimensionless code inputs $\bar\mu_b=0.1$, $\bar\eta=0.01$, and $\bar\lambda=1$, which correspond in the code units to $\mu_b\simeq10^{-16}\,\mathrm{MeV}$, $\eta=1$, and $\lambda=100$. These values determine the bosonic mass scale $Z_b=\lambda^{-1/2}\mu_b^2$ and, together with the AkmalPR EOS, allow the final results to be expressed in physical units, while also placing the bosonic sector in the strong self-interaction regime required by the RPV approximation \cite{Ryan:1996nk,Vaglio:2022flq,Mourelle:2024qgo}. Moreover, for a boson star one always has $\bar{w} \in [0, \bar{\mu}_b]$.

The numerical integration of the equilibrium equations follows the same strategy adopted in \cite{Mourelle:2024qgo}, namely a modified version of the \texttt{RNS} code based on the KEH self-consistent field method \cite{komatsu1989rapidly,friedman2013rotating}. The equilibrium models are constructed through an iterative procedure in which the matter distribution and the metric potentials are updated successively until the sources obtained from the matter variables are consistent with the spacetime geometry they generate.

This approach solves the reduced field equations by using Green's functions. Those equations are explicitely reported in Appendix~\ref{appx_source}. In Eqs.~\eqref{eq:DiffEq_gamma}-\eqref{eq:DiffEq_alpha}, each equation is arranged so that the left-hand side contains a linear differential operator acting on the corresponding metric function. This operator contains the second-order derivative structure of the equation, together with the lower-order derivative terms required by the chosen coordinates. The remaining terms, including the nonlinear metric contributions and the matter sources, are collected on the right-hand side and treated as effective sources. The Green function used for each equation is the fundamental solution of the corresponding linear operator, i.e. the solution associated with a localized source and satisfying the homogeneous equation away from that source. In this sense, the Green function encodes the solution of the left-hand-side operator with the appropriate regularity and asymptotic behaviour. Therefore, the boundary conditions are not imposed separately at each iteration; rather, regularity at the origin and asymptotic flatness at infinity are automatically incorporated through the choice of Green functions. Starting from an initial equilibrium configuration, the effective sources are computed from the current matter distribution and metric functions. Those are then updated using the corresponding Green-function solutions, and the matter variables are reconstructed consistently with the new geometry and rotational contributions. This procedure is iterated until the spacetime geometry and matter distribution form a self-consistent solution of the coupled system.

The functions to be solved in the code are written in terms of a compactified radial coordinate
\begin{equation}
r = r_{e}
\frac{s}{1-s},
\qquad
0\leq s \leq 1\,.
\end{equation}
The equatorial radius $r_e$ is related to the polar radius $r_p$ through the axis ratio $r_{\mathrm{ratio}}=r_p/r_e$. We also introduce the angular variable
 $   \mu = \cos\theta $ \footnote{It should
not be confused with \(\mu_b\), which denotes the boson mass.},
used only as a redefinition of the polar coordinate. Moreover, the circumferential radius $\mathcal{R}$ in the physical (Jordan) frame, is defined as
\begin{equation}
\mathcal{R} = A(\varphi)\,r\,
\exp\!\left[\frac{\gamma-\sigma}{2}\right].
\label{eq:R_circ}
\end{equation}
The equatorial circumferential radius
of the fermionic core is denoted by $R_e$ and defined as the value of
$\mathcal{R}$ at the fermionic surface for $\theta=\pi/2$.

The three elliptic metric potentials, $\rho$, $\gamma$, and $\sigma$, together with the scalar field contribution $\varphi$ are obtained by expanding the Green’s functions into their radial and angular parts. The source terms for $R$-squared gravity
\begin{equation}
\begin{split}
&\Hat{\Hat{S}}_T^\rho(s,\mu)=\bar{r}^2 S_T^\rho(s,\mu), \\
&
\Hat{\Hat{S}}_T^\gamma(s,\mu)=\bar{r}^2 S_T^\gamma(s,\mu), \\
&
\Hat{\Hat{S}}_T^{\sigma}(s,\mu)=\bar{r}_e \bar{r}^2 S_T^\sigma(s,\mu), \\
&
\Hat{\Hat{S}}_T^{\varphi}(s,\mu)=\bar{r}^2 S_T^\varphi(s,\mu),
\end{split}
\end{equation}
are then integrated using Legendre and associated Legendre polynomials, $P_n(\mu)$ and $P_n^m(\mu)$. Note that the source terms $S_T^x(s,\mu)$ contain radial and angular derivatives of the metric potentials\footnote{The explicit form of the source terms  for $R$-squared gravity is reported in Appendix \ref{appx_source}.}. An important point in our implementation of the equations is that the modified-gravity contribution is treated on the source terms. This means that just as in the case of the complex bosonic field, the terms associated with the effective scalar field $\varphi$ enter the code as  matter-like source terms. 

By contrast, the equation for the metric potential $\alpha$ is not elliptic, but rather a first-order ordinary differential equation, and is therefore obtained directly by integrating its derivative expression. It is also worth emphasizing that the rescalings in terms of $r_e$ follow the same prescription as in \cite{komatsu1989rapidly,friedman2013rotating, Mourelle:2024qgo}, where both rotating neutron stars and rotating fermion-boson stars are treated consistently. Within this framework, $r_e$ continues to play the role of the characteristic radial scale entering the compactified coordinate mapping and the associated dimensionless variables. 

A key part of our numerical procedure is the construction of the initial guess used in subsequent integrations to build the rotating models. For this purpose, we first solve the purely static NS configuration (see Appendix~\ref{app:static_spacetime} for details). Since the bosonic field is incorporated through the RPV approximation, the code does not require an independent spherical purely bosonic solution as an initial guess. Instead, the bosonic field can be determined from its own parameters together with the metric functions, as shown in Eq.~\eqref{ryan_field}. Therefore, once the static NS solution has been obtained through a standard Runge-Kutta integration, the corresponding non-rotating metric functions are used as the seed configuration, from which an initial profile for $\phi$ is constructed and employed as the starting point for the iterative rotational scheme. 

\section{Global properties}\label{sec:iv}
\label{section4}

The global properties of rotating fermion-boson stars in $R$-squared gravity can be determined from the asymptotic behavior of the spacetime. In stationary and axisymmetric spacetimes, the gravitational mass and angular momentum are obtained through the Komar integrals constructed from the timelike and azimuthal Killing vectors $\xi^{\beta}_{(t)}$ and $\xi^{\beta}_{(\psi)}$, 
\begin{equation}
\begin{split}
M &= -\int \left(2{T}^{\ast \alpha}_{\beta}
-\delta^{\alpha}_{\beta}{T}^{\ast \rho}_{\rho}\right)
\xi^{\beta}_{(t)} d^3\rho_{\alpha}
\\
&=
\int
\left(
-2{T}^{\ast t}_{t}
+{T}^{\ast \rho}_{\rho}
\right)
\sqrt{-{g}^{\ast}}\, d^3x ,
\\
J &=
\int
{T}^{\ast \alpha}_{\beta}
\xi^{\beta}_{(\psi)}
d^3\rho_{\alpha}
=
\int
{T}^{\ast t}_{\psi}
\sqrt{-{g}^{\ast}}\, d^3x .
\end{split}
\end{equation}
The dimensionless gravitational mass associated with the fermionic sector is given by
\begin{eqnarray}
\bar{M}_{f}
&=&
4\pi \bar{r}_e^3
\int_0^1
\frac{s^2 ds}{(1-s)^4}
\int_0^1
d\mu \,
e^{2\alpha+\gamma}
\nonumber\\
&&\times
\Bigg[
A^4(\varphi)
\left(
\frac{\bar{p}_{f}
+\bar{\rho}_{f}}
{1-v^2}
\right)
\Bigg(
1+v^2
\nonumber\\
&&\qquad
+
\frac{2vs}{1-s}
\sqrt{1-\mu^2}\,
e^{-\rho}\bar{\sigma}
\Bigg)
+
2A^4(\varphi)\bar{p}_{f}
\nonumber\\
&&\qquad
-
\frac{1}{8\pi}
V(\varphi)
\Bigg],
\end{eqnarray}
while the fermionic contribution to the dimensionless angular momentum reads
\begin{eqnarray}
\bar{J}_{f}
&=&
4\pi \bar{r}_e^4
\int_0^1
\frac{s^3 ds}{(1-s)^5}
\int_0^1
d\mu \,
\sqrt{1-\mu^2}
\nonumber\\
&&\times
e^{2\alpha+\gamma-\rho}
A^4(\varphi)
\frac{
(\bar{p}_{f}
+\bar{\rho}_{f})v
}
{1-v^2}.
\end{eqnarray}

The bosonic contribution to the dimensionless mass is obtained from the rotating complex field configuration,
\begin{eqnarray}
\bar{M}_{b}
&=&
4\pi \bar{r}_e
\int_0^1
\frac{ds}{(1-s)^2}
\int_0^1
d\mu \,
e^{2\alpha-\rho}
|\Phi|^2
A^4(\varphi)
\nonumber\\
&&\times
\Bigg[
e^{2\rho}
\frac{\bar{\eta}^2}{1-\mu^2}
+\frac{s^2}{(1-s)^2}
\Bigg(
\bar{w}^2
-\bar{\eta}^2\bar{\sigma}^2
\nonumber\\
&&\qquad
+
e^{\gamma+\rho}
\frac{\bar{\lambda}}{2}
|\Phi|^2
\Bigg)
\Bigg],
\end{eqnarray}
whereas the bosonic contribution to the dimensionless angular momentum is
\begin{eqnarray}
\bar{J}_{b}
&=&
4\pi \bar{r}_e^3
\int_0^1
\frac{s^2 ds}{(1-s)^4}
\int_0^1
d\mu \,
e^{2\alpha-\rho}
A^4(\varphi)
\nonumber\\
&&\times
\bar{\eta}
(\bar{w}-\bar{\eta}\bar{\sigma})
|\Phi|^2 .
\end{eqnarray}
The masses and angular momenta quoted in this section are the ADM quantities measured by a distant observer. Since the scalar field vanishes at infinity, $\varphi\to 0$ and $A(\varphi)\to 1$, the Jordan-frame and Einstein-frame ADM masses and angular momenta coincide in $R$-squared gravity \cite{Doneva:2013qva}. The same statement applies to the total mass $M_T$, as well as to its fermionic and bosonic components. The total angular momentum $J_T$ is defined in the usual way and is identical in both frames. The equatorial circumferential radius $R_e$ is likewise a physical, Jordan-frame quantity, defined through the circumferential radius $\mathcal{R}$ in Eq.~\eqref{eq:R_circ}. By contrast, $\bar w$ and $\bar\eta$ are dimensionless parameters characterizing the bosonic field configuration rather than asymptotic observables, and therefore they are not subject to this frame comparison.

The quantities $\bar{w}$ and $\bar{\eta}$ denote the dimensionless boson frequency and rotational quantum number, respectively, while $\bar{\sigma}$ represents the frame dragging angular velocity. The conformal factor $A^4$ modifies both the fermionic and bosonic sectors through the coupling between matter and the scalar degree of freedom associated with the quadratic curvature correction. In the numerical implementation, these integrals are evaluated using Simpson integration over the compactified radial coordinate $s$ and the angular coordinate $\mu$. The source terms entering the mass and angular momentum integrals therefore differ from the GR case (see~\cite{Mourelle:2024qgo}) through the appearance of the conformal coupling factor and the scalar potential contribution. The total dimensionless mass and angular momentum can be decomposed into fermionic and bosonic contributions
\begin{equation}
\bar{M}_{T}=\bar{M}_{f}+\bar{M}_{b},
\qquad
\bar{J}_{T}=\bar{J}_{f}+\bar{J}_{b}.
\end{equation}
The scalar degree of freedom contributes implicitly through the modified metric functions, the conformal coupling factor $A(\varphi)$, and the scalar potential appearing in the gravitational field equations. 

\section{Results}\label{sec:v}

Our main goal is to construct rotating fermion-boson stars within the framework of $R$-squared gravity for different values of $a$ in Eq.~(\ref{f_of_r}), the parameter that controls the deviation from GR. For completeness, we also provide the solutions in the nonrotating (static) limit. For both the static and rotating cases, we consider three distinct situations, namely pure GR together with two representative cases of $R$-squared gravity, $a=10$ and $a=10^4$. We recall that varying $a$ changes the effective mass scale of the scalar field and, consequently, affects the equilibrium structure of the solutions.

The discussion is organized in three subsections: a first one devoted to the static limit, a second one addressing rotating configurations, and a third one focusing on mass-radius relations for both static models and configurations at the mass-shedding limit. The static and Keplerian limits define the boundaries of the equilibrium sequences which allows us to assess the consistency of the mass-radius relations with current astrophysical constraints. Additional models are discussed in appendices~\ref{appendix0} and~\ref{appendixb}.

\subsection{Static limit}

The static limit of mixed compact stars in $R$-squared gravity has been studied previously in \cite{2018PhRvD..97b4030L}, where spherically symmetric configurations with an admixed dark component were considered in the Starobinsky model. Our static results are qualitatively consistent with that work: larger values of $a$ increase the maximum supported mass and modify the internal distribution of both matter sectors relative to GR. Quantitative differences are expected because we use the AkmalPR EOS and treat the bosonic sector within the RPV approximation. 

We obtain the static configurations as the $J_T \to 0.0 M_{\odot}^2$ limit of our rotating numerical scheme. In this limit, using the equations reported in Appendix~\ref{app:static_spacetime}  as a reference, all azimuthal dependence vanishes and the $\theta$-dependence becomes trivial, so that the fermionic energy density, bosonic field profile, and metric functions depend only on the radial coordinate $r$. For instance
\begin{equation}
    \rho_f(r,\theta) \longrightarrow \rho_f(r),
    \qquad
    \Phi_{s}(r,\theta) \longrightarrow \phi_{s}(r).
\end{equation}

\begin{figure*}[t]
\subfloat[]{
\includegraphics[width=\textwidth]{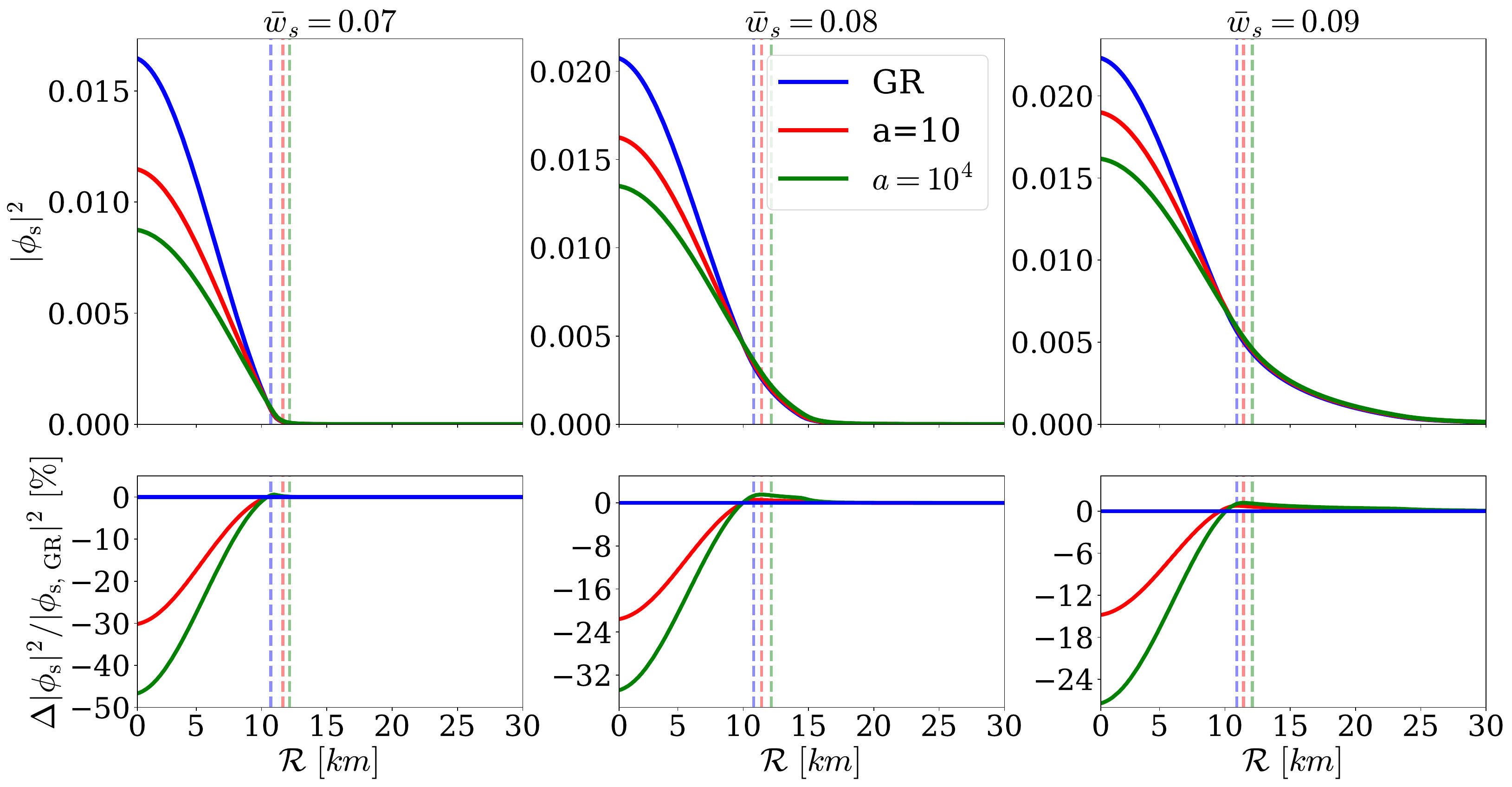}
\label{fig1a}
}
\vspace{0.5cm}
\subfloat[]{
\includegraphics[width=\textwidth]{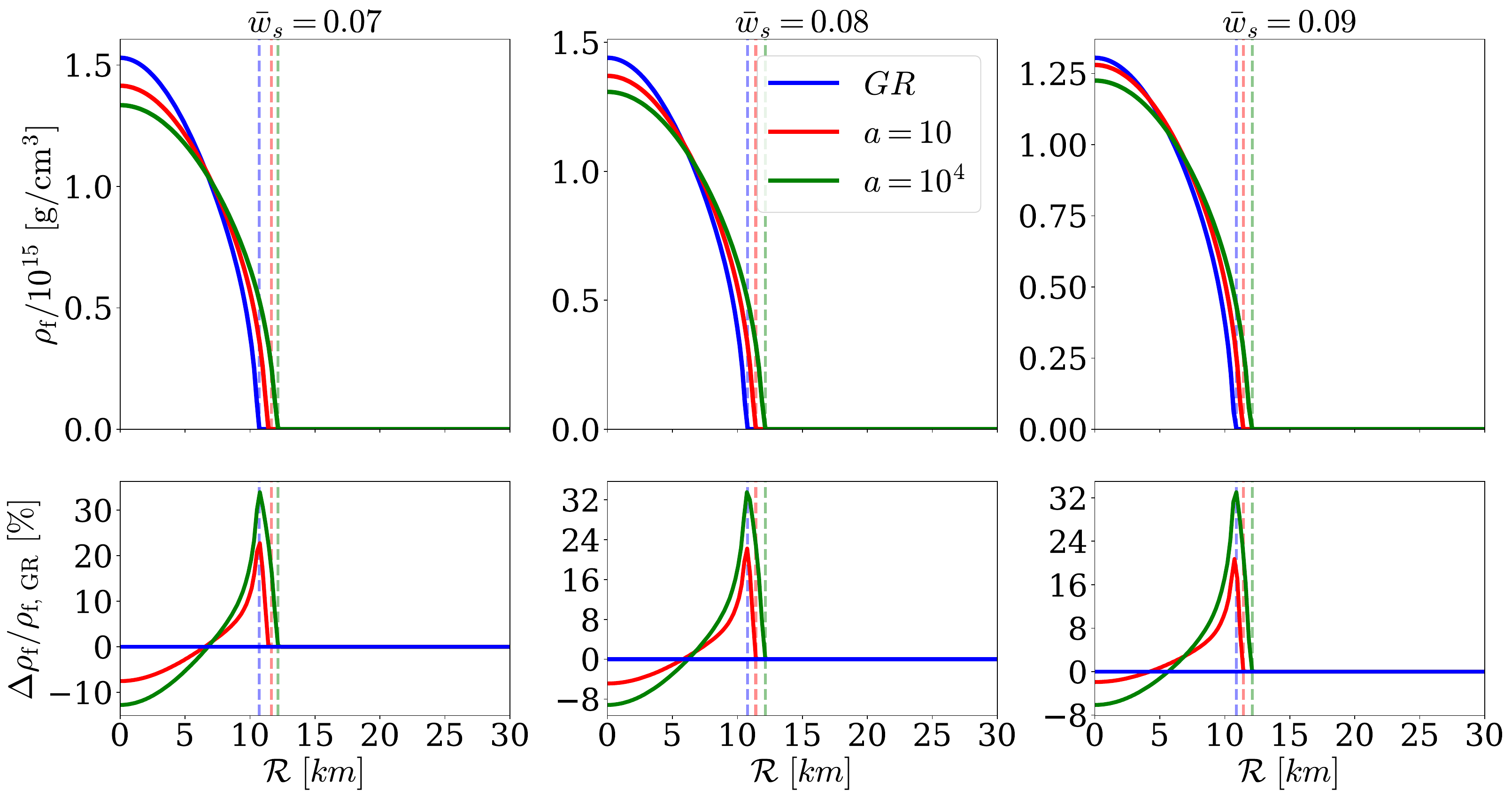}
\label{fig1b}
}
\caption{Radial profiles of the  bosonic field ({\bf Top a}) and  fermionic energy density ({\bf Top b}) for static mixed star configurations in GR and two different models of $R$-squared gravity, with $J_{T}=0.0 M_{\odot}^2$, $M_{T}=2.0\,M_{\odot}$, and three different values of $\bar{w}$. The dashed lines indicate the fermionic radii $R_e$ for each theory. {\bf Bottom a and b}: Corresponding relative percentage difference for the $R$-squared gravity models compared to the corresponding results obtained within GR. 
}
\label{fig1}
\end{figure*}

For the fermionic energy density, the static limit emerges in a rather direct manner, since the overall topology of the matter distribution remains qualitatively unchanged between rotating and nonrotating configurations. The bosonic sector, however, exhibits a more nontrivial behavior. Static boson star solutions possess a spherically symmetric topology, characterized by a regular field profile that attains its maximum value at the origin and decays monotonically with radius. In contrast, rotating bosonic configurations develop a toroidal structure associated with a nonzero azimuthal harmonic index. 

In the RPV framework the bosonic field is treated within a high-coupling approximation, avoiding the need to solve an eigenvalue problem for the $w_s$, as detailed above. Within this setting, one expects a smooth deformation of the solutions as the harmonic index is reduced,  with rotating configurations continuously approaching spherical (or quasi-spherical) profiles in the limit of vanishing azimuthal number. 

Fig.~\ref{fig1} shows the radial profiles of the static fermion-boson stars obtained in GR and in the two representative models of $R$-squared gravity, for a fixed total gravitational mass $M_{T}=2.0\,M_{\odot}$ and vanishing total angular momentum $J_{T}=0.0 M_{\odot}^2$.  The upper panels of Fig.~\ref{fig1} (a) display the modulus squared of the bosonic field as a function of $\mathcal{R}$, while the upper panels of Fig.~\ref{fig1} (b) show the corresponding fermionic energy density. Each row is organised into three columns corresponding to the three chosen values of the dimensionless bosonic internal frequency, which constitutes the primary control parameter of the bosonic sector in the static limit.  The vertical dashed lines denote the equatorial circumferential radius of the fermionic core, defined as the stellar radius where the fermionic energy density vanishes. This radius serves as a natural reference scale for assessing the spatial extent of the bosonic field.  Also, in both figures, the lower sub-panels quantify the relative percentage deviation between each $R$-squared gravity model and the corresponding GR solution.  

The role of the bosonic internal frequency $\bar{w}_{s}$ on the spatial structure of the bosonic part is evident from the left-to-right comparison across the three columns. For the lowest frequency considered, $\bar{w}_{s}=0.07$, the bosonic field extends over approximately the same spatial region as the fermionic matter. The profile of $|\phi_{s}|^{2}$ attains its maximum at the stellar centre and decreases rapidly inside the fermionic core, indicating that the two matter components are effectively co-spatial. In this regime, the dark bosonic component is embedded within the ordinary NS matter. Although not shown here, more extreme configurations with lower values of $\bar{w}_{s}$ lead to an even more compact bosonic distribution, which can become fully embedded within the fermionic component or even confined to a core-like region. As $\bar{w}_{s}$ is increased, the bosonic-field envelope broadens, and its support extends well beyond the fermionic surface, giving rise to a bosonic halo surrounding the NS.  At $\bar{w}_{s}=0.09$ in particular, the tail of $|\phi_{s}|^{2}$ (and thus a non-negligible amount of bosonic matter) extends to radii well exceeding the outer boundary of the fermionic region. 

The modifications induced by $R$-squared gravity on the bosonic field profile are significant across all three frequency values, as illustrated in the lower subpanels of Fig.~\ref{fig1} (a). Relative to the GR case, both modified-gravity models lead to a suppression of $|\phi_s|^2$, with the effect being most pronounced in the inner region of the configuration and gradually decreasing toward the outer layers. As the radius increases, the profiles exhibit a crossing point, appearing at lower radii for models with higher field frequencies. Hence, although the $R$-squared gravity configurations have lower central amplitudes, their scalar-field distributions extend to larger radii and vanish farther from the center. As $\bar{w}_s$ increases, the relative deviation between the GR and $a=10$ cases decreases, while the difference between the $a=10$ and $a=10^4$ models is likewise reduced. This suggests a progressive saturation of the modified-gravity effects at higher bosonic field frequencies.

The fermionic energy-density profiles displayed in the upper panels of Fig.~\ref{fig1} (b), together with the associated percentage deviations shown in the lower subpanels, exhibit a qualitatively distinct response to the modified-gravity corrections.
As $a$ increases, the central energy density $\rho_f(0)$ decreases. Relative to the GR configuration, the $a=10^{4}$ case shows the most spatially extended fermionic matter
distribution, despite the total mass remaining fixed. This should be interpreted not as a depletion of the fermionic content, but
rather as a redistribution of matter over a moderately enlarged stellar volume. The value $a=10$ represents a moderate deviation from GR, while $a=10^4$ corresponds to a very light scalar field approaching the maximum deviation from GR within this model, consistent with previous studies \cite{Yazadjiev:2015zia,Doneva:2015hsa}.

When considering the fermionic sector, the percentage-deviation profiles indicate that, in both modified-gravity models, departures from the GR solution remain small both near the stellar center and in the outermost layers. A localized positive enhancement, however, emerges at intermediate radii within the fermionic core. This reflects the influence of
the scalar degree of freedom on the internal pressure-support balance: the modified
gravitational interaction favors a more extended fermionic
configuration, effectively shifting matter from the central region
toward the outer layers. This indicates that the modified-gravity effects are strongest when both sectors are spatially co-located and weaken as the bosonic halo tends to move outward relative to the fermionic core.

\subsection{Rotating solutions}

We turn now to analyze the rotating configurations following a procedure analogous to that used for the static case. For the same three gravitational models introduced above, we compute equilibrium solutions for various values of the dimentionless bosonic internal frequency $\bar{w}$, while keeping fixed the global quantities $J_{T}=1.0 M_{\odot}^2$ and $M_{T}=2.0 M_{\odot}$ (which we call Set 1), and also for $J_{T}=2.0 M_{\odot}^2$ and $M_{T}=2.0 M_{\odot}$ (Set 2; these are discussed in Appendix~\ref{appendix0}). For each choice of $\bar{w}$, $J_{T}$ and $M_{T}$, the code adjusts the central energy density and the angular velocity until a consistent equilibrium solution is obtained. 

\begin{figure*}[ht]
 \centering
 \includegraphics[angle=0,width=1\hsize]{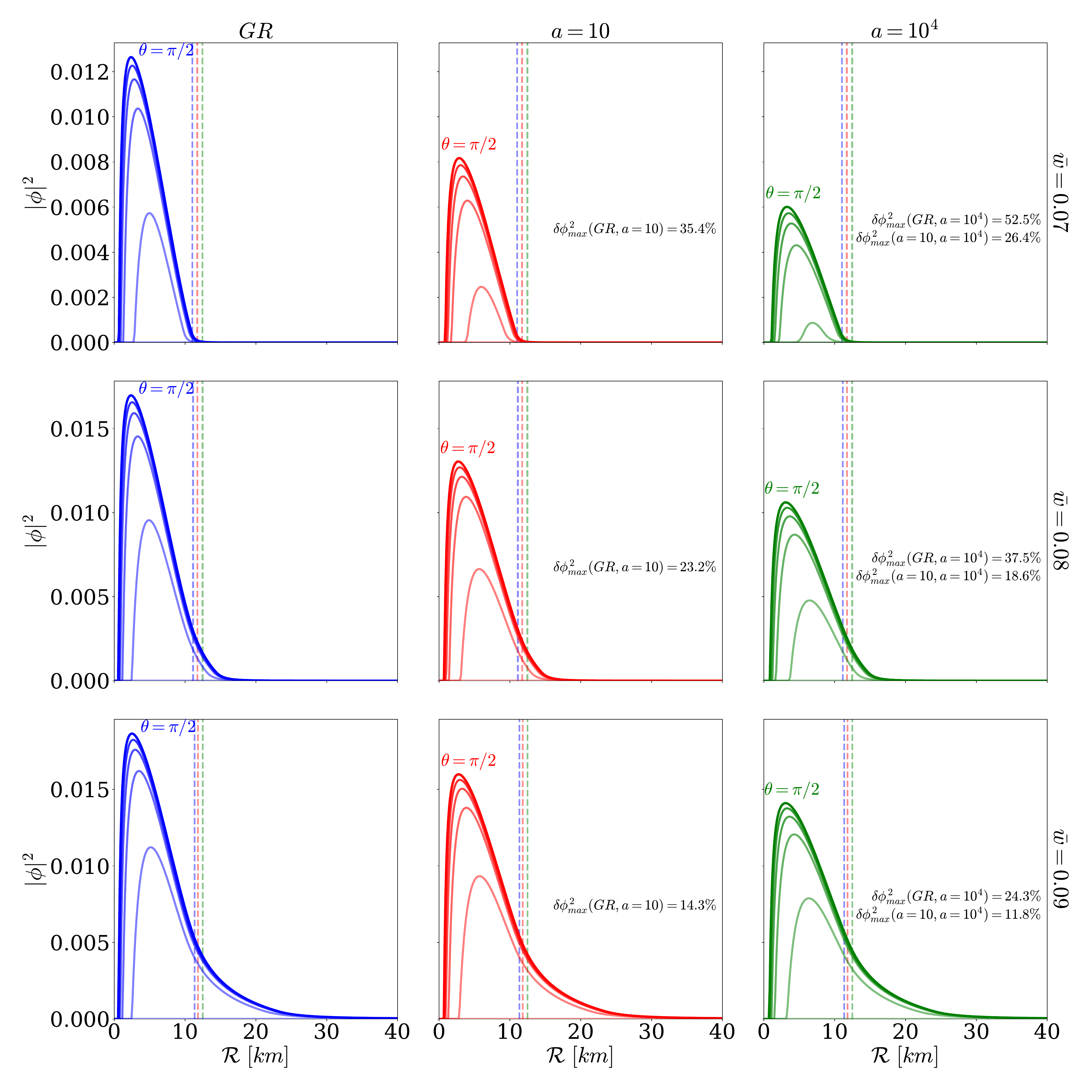}
 \caption{Radial profiles of $|\phi|^2$ for rotating mixed-star models in GR and two $R$-squared gravity cases ($a=10$ and $a=10^4$), with $J_T=1.0 M_{\odot}^2$, $M_T=2.0\,M_\odot$, and three values of $\bar w$. In each panel, the solid curves show $|\phi|^2$ for a discreet set of polar angles $\theta=\pi/12, \pi/6, \pi/4, \pi/3$, and $\pi/2$ (note that the bosonic field vanishes on the rotation axis $\theta=0$). The upper boundary corresponds to the equatorial plane ($\theta=\pi/2$), where $|\phi|^2$ is maximal. The curves become increasingly crowded toward the equator because the bosonic distribution is concentrated there. Vertical dashed lines mark the equatorial circumferential radius of the fermionic core in each theory. }
\label{fig2}
\end{figure*}

Fig.~\ref{fig2} shows $|\phi|^2$ as a function of the circumferential radius $\mathcal{R}$ for the rotating configurations of Set 1. Each panel shows a selection of radial profiles generated by sampling the polar angle at discrete values within the range $0\leq \theta \leq \pi/2$. The lower edge of the profile bundle is the rotation-axis profile ($\theta=0$). For rotating boson star solutions this curve lies on the horizontal axis, since $|\phi(\mathcal{R},\theta=0)|=0$ everywhere on the axis. The upper edge is the equatorial profile ($\theta=\pi/2$), which gives the largest values of $|\phi|^2$ at each radius. The intermediate curves fill the region between these two limits and become progressively denser near the equator, where the bosonic matter is most concentrated. The vertical dashed lines indicate the fermionic equatorial radius, evaluated at $\theta=\pi/2$, in each gravitational theory. The rotating configurations differ from the static limit in two key aspects. First, the bosonic field vanishes at the coordinate origin, $\phi(0)=0$, leading to a toroidal matter distribution that is characteristic of rotating boson stars. Second, the field acquires a non-trivial dependence on the polar angle $\theta$, reflecting the rotational departure from spherical symmetry. Specifically, the bosonic field varies throughout the interior as a function of both $\mathcal{R}$ and $\theta$, vanishing along the rotation axis and reaching its maximum value at the equatorial plane, $\phi(\mathcal{R}, \theta=\pi/2)=\phi_{\max}$.

\begin{figure*}[!ht]
 \centering
  \includegraphics[angle=0,width=1\hsize]{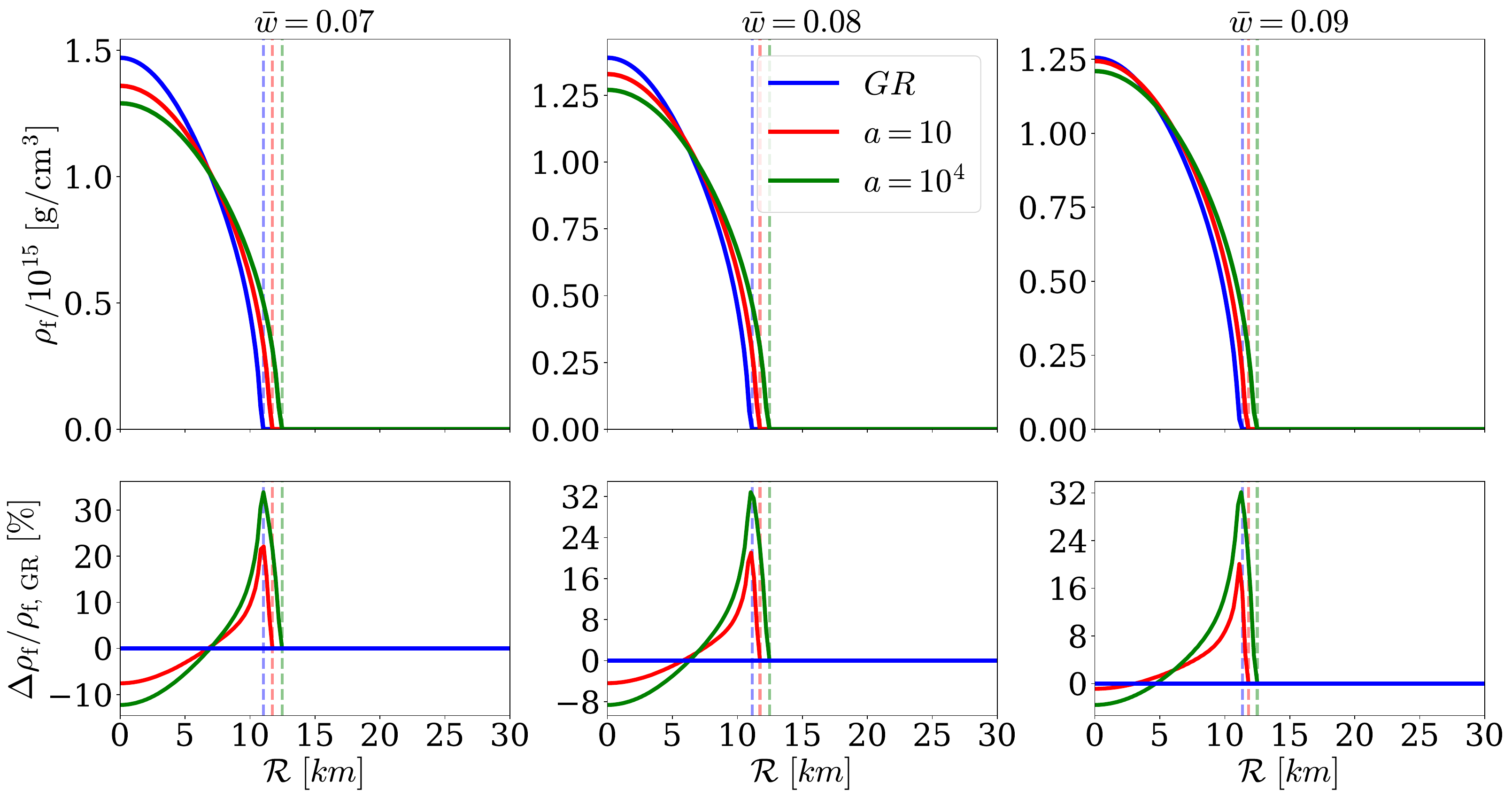}
 \caption{{\bf Top}: Radial profiles of the fermionic energy density $\rho_f$ for rotating mixed-star configurations in GR and two representative $R^2$ gravity, for $J_T = 1.0\,M_{\odot}^2$ and $M_T = 2.0\,M_\odot$ (Set 1), and three values of $\bar{w}$ at the equatorial plane ($\theta = \pi/2$). {\bf Bottom}: Relative percentage deviation of $\rho_f$ from the corresponding GR solution. Dashed lines denote the location of the fermionic surface in each gravitational theory.} 
\label{fig3}
\end{figure*}

The dimensionless internal bosonic frequency $\bar{w}$ acts as the primary control parameter governing the spatial extent and compactness of the bosonic sector. Smaller values of $\bar{w}$ correspond to more compact configurations, whereas larger values produce increasingly dilute, spatially extended distributions for a fixed total mass. This determines whether the bosonic component remains confined within the stellar interior or extends outward to form a halo-like structure surrounding the fermionic counterpart.

The results show that the presence of the scalar degree of freedom can lead to quantitative modifications to the bosonic profile. Comparing the columns in Fig.~\ref{fig2}, one can observe a systematic suppression of the maximum value of $|\phi|^2$ in the $R$-squared gravity models relative to the GR baseline. Specifically, the peak value of the bosonic field in the GR case is larger than that found in the corresponding $a=10$, and $a=10^4$ configurations depending on the specific frequency $\bar{w}$. Moreover, the choice of gravitational model also influences the angular distribution of the bosonic field. The transition from GR to the $a=10$ case produces a substantial structural change in the bosonic field profile, whereas further increasing the parameter to $a=10^4$ leads to smaller incremental corrections.

Fig.~\ref{fig3} displays the corresponding fermionic energy density profiles, $\rho_f$, for the same rotating configurations of Set 1 evaluated at the equatorial plane, $\theta = \pi/2$. While the global structure of the fermionic sector is less sensitive to the model parameters than the bosonic one, the scalar degree of freedom still induces a non-trivial redistribution of the baryonic matter. In both modified-gravity scenarios, the central energy density $\bar{\rho}_f(0)$ systematically decreases relative to the GR baseline. This is accompanied by a localized positive enhancement of the energy density at intermediate equatorial radii within the fermionic core, as illustrated by the relative percentage deviations in the bottom panels of Fig.~\ref{fig3}. Similarly to the static case, the addition of $R$-squared contributions in the rotating configurations change the internal balance between gravity and pressure support, thereby favoring a more spatially extended fermionic distribution. In addition, as $\bar{w}$ increases and the bosonic field develops a halo-like structure, the spatial overlap between the fermionic and bosonic sectors gradually decreases. Hence, the scalar degree of freedom-induced modifications to $\rho_f$ are most pronounced when both components are typically distributed over similar spatial domains.

The space of equilibrium fermion-boson star solutions is enlarged when adding $R$-squared terms and progressively expands as $a$ increases. In particular, reproducing the same total mass and angular momentum in theories with larger values of $a$ requires smaller central or peak values of the matter variables, reflecting a modified effective gravitational coupling that reduces the localized matter concentration needed for a given mass budget. By contrast, when the central fermionic density and bosonic frequency are allowed to vary freely, the enlarged solution space extends to higher limiting masses and broader admissible mass-radius combinations than in GR. Whether this also leads to systematically different configurations will be examined in the following subsection through the analysis of the mass-radius relations. 

The structural properties of the configurations in Set 2 are discussed in Appendix~\ref{appendix0} (see Figs.~\ref{fig5} and \ref{fig6}). For this set of models, the fixed angular momentum is twice that of Set 1. The qualitative trends found for Set 2 are highly consistent with those observed in Set 1. Specifically, the response of the matter distributions to variations in the modified-gravity parameter $a$ remains qualitatively similar. While the relative variations in the maximum amplitude of the bosonic field remain comparable between both sets, the corresponding modifications to the fermionic energy density are notably more suppressed in Set 2. This indicates that under rapidly rotating regimes, the centrifugal effects partially dominate the structural equilibrium, rendering the fermionic distribution less sensitive to the scalar field than in slower rotation regimes. 

We close this section by noting that while the primary configurations analyzed in this section do not represent core-like bosonic distributions, such topologies are accessible within the broader parameter space of our models. In particular such configurations  emerge in regimes characterized by a sufficiently large bosonic mass fraction and a sufficiently low internal frequency (typically $\bar{w} < 0.06$ for high-mass models). This is shown in Appendix~\ref{appendixb} (in particular, see the yellow regions in the color maps of Fig.~\ref{fig_fin}). Furthermore, increasing the value of $a$ systematically enlarges this region of the parameter space, as evidenced by the emergence of new high-mass, low-$\bar{w}$ equilibrium configurations in the color maps. This trend suggests that increasing $a$ enhances the impact of the scalar-field sector on the gravitational equilibrium. In practice, this allows configurations to exist in parts of the parameter space which lead to a more compact and more concentrated bosonic sector, in the inner region of the star. Therefore, larger values of $a$ tend to favor core-like bosonic distributions, particularly for high-mass models with low values of $\bar{w}$.

\subsection{Mass-radius relations and observational bounds}

We now present a set of mass-radius relations for our three gravitational models (GR, $a=10$, and $a=10^4$). These diagrams provide a convenient framework for illustrating the influence of the $R$-squared modifications of GR, and for assessing how the coupling parameter $a$ impacts the global observable properties of the equilibrium configurations. In particular, they allow us to examine how the underlying gravitational theory modifies the admissible ranges of masses and radii, thereby affecting the overall structure and stability domain of the solutions.

\begin{figure*}[t]
\centering
 \includegraphics[width=\textwidth]{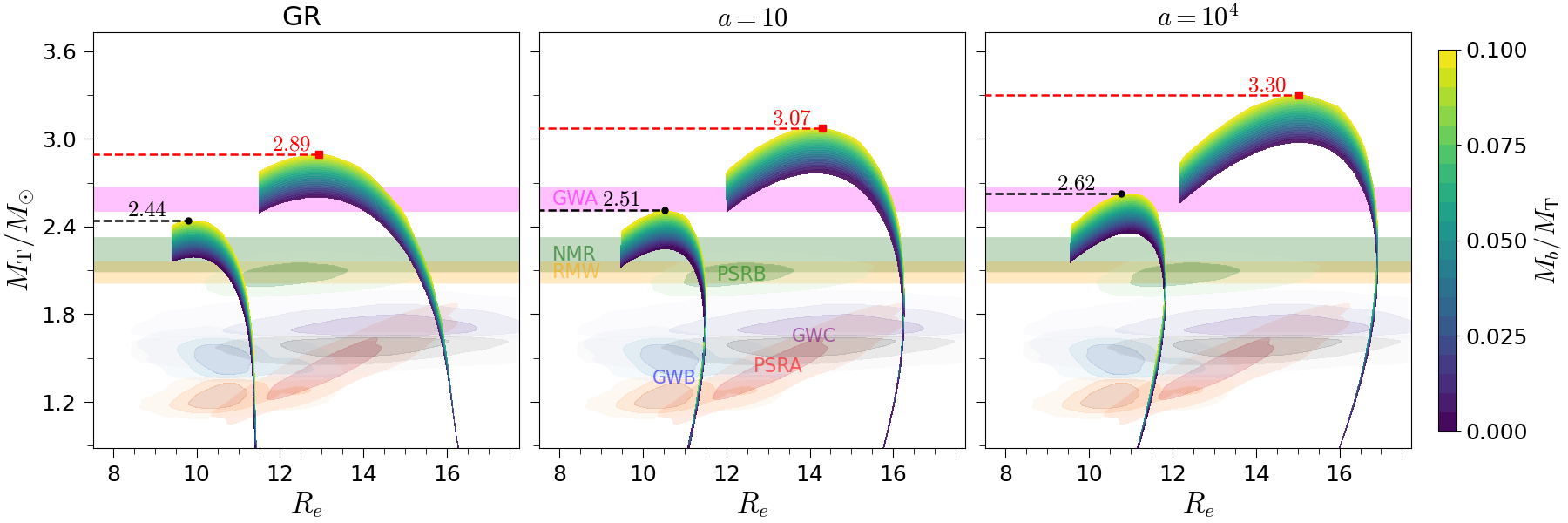}
 \caption{Mass-radius relations for mixed star configurations in the framework of GR and two distinct models of $R$-squared gravity, shown for the two critical regimes of static equilibrium (contour plots on the left side of each panel) and rotation at the Keplerian frequency (contours on the right side of each panel). The different colors within the colormap represent the fractional contribution of the bosonic mass to the total gravitational mass. The AkmalPR EOS is considered. The observational and theoretical constraints shown in the figure are
labelled as follows: GWA denotes GW190814~\cite{LIGOScientific:2020zkf}, GWB
denotes GW170817~\cite{LIGOScientific:2017vwq}, GWC denotes
GW190425~\cite{LIGOScientific:2020aai}, NMR refers to Nathanail
et al.~(2021)~\cite{Nathanail:2021tay}, RMW to Rezzolla
et al.~(2018)~\cite{Rezzolla:2017aly}, PSRA to PSR~J0030+0451
\cite{Riley:2019yda}, and PSRB to PSR~J0740+6620
\cite{Legred:2021hdx}. }
 \label{fig9}
\end{figure*}

Fig.~\ref{fig9} displays mass-radius diagrams for static and rotating fermion-boson star configurations, with each panel corresponding to a distinct gravitational
theory. The sequences have been built using the tabulated AkmalPR EOS for the fermionic component. The horizontal axis is the equatorial circumferential radius, $R_e$, while the vertical axis shows the total gravitational mass, $M_T$. For each theory, two families of equilibrium solutions are shown: the static sequences, which populate the contour region on the left side of each panel, and the configurations at the Keplerian, or mass-shedding, limit, which populate the region on the right side. Together,
these two families delimit the complete space of physically admissible
rotating equilibria: any configuration with uniformly rotating fermionic sector must reside between the static and the Keplerian boundary in the mass-radius plane.  Rather than tracing
individual one-dimensional curves, the solution space is represented
as a two-dimensional filled region. The reason is because, for preparing these plots, both the central fermionic energy density and the bosonic internal frequency have been allowed to vary freely while the bosonic mass fraction has been constrained to remain at or below $10\%$ of the total gravitational mass. This upper limit reflects the expectation that, in scenarios in which the bosonic component is gravitationally captured or accreted onto an already formed NS, the fermionic matter remains dominant and the accumulated dark component contributes only a small fraction of the total mass.
Static and dynamical studies of fermion-boson stars find boson-to-fermion 
number ratios of order $10\%$ or smaller in such configurations 
\cite{HENRIQUES198999,Henriques:1989ez,Henriques:1990xg,DiGiovanni:2020frc,DiGiovanni:2021ejn}, 
and similar conclusions hold for dark-matter-admixed NSs \cite{Cipriani:2025tga}. 

The continuous color map in Fig.~\ref{fig9} encodes this bosonic fraction for each equilibrium solution.  The static maximum mass of each sequence is indicated by a horizontal black dashed line, while the corresponding Keplerian maximum mass is marked by a horizontal red dashed line, with numerical values annotated directly in each panel to facilitate precise cross-theory comparison. Moreover, several observational and theoretical constraints are superimposed on each panel and will be discussed below in the context of the astrophysical viability and consistency of our models. These constraints provide a useful framework for confronting the predictions of our analysis with current astrophysical measurements, gravitational-wave observations, and theoretical bounds associated with compact stellar configurations.

The most apparent result from Fig.~\ref{fig9} is the
systematic increase of both the static and Keplerian maximum masses
as the $R$-squared gravity parameter $a$ is increased.  In the GR case,
the maximum mass of the static sequence reaches $2.44\,M_{\odot}$,
while the Keplerian limit extends this to $2.89\,M_{\odot}$.
Introducing the scalar degree of freedom with $a=10$ raises these values to
$2.51\,M_{\odot}$ and $3.07\,M_{\odot}$, corresponding to
percentage increases of $2.87\%$ and $6.18\%$ relative to GR for the
static and Keplerian cases respectively.
Increasing the mass scale of the scalar field further to $a=10^{4}$ amplifies
these enhancements substantially: the static maximum mass grows to
$2.62\,M_{\odot}$ ($+7.47\%$ relative to GR) and the Keplerian
maximum mass reaches $3.29\,M_{\odot}$ ($+13.94\%$ relative to GR). It is worth stressing that the enhancement of the maximum mass is consistently larger in the Keplerian limit than in the static limit for every value of $a$. This comparison refers to the boundaries of the equilibrium sequences and should not be confused with the profile analysis of the previous subsection, where $M_T$, $J_T$, and $\bar{w}$ were held fixed: in that case, a larger angular momentum partially suppresses the modified-gravity-induced changes in the central fermionic energy density. 

The two-dimensional character of the solution space in 
Fig.~\ref{fig9}, encoded by the color map of the bosonic mass fraction, 
reveals a rich internal structure that is absent from single-fluid NS 
mass-radius diagrams. Within the RPV approximation, the bosonic field profile is controlled by the 
central fermionic energy density and the internal field frequency $\bar{w}$, 
with the amplitude determined algebraically by Eq.~\eqref{ryan_field}. 
For a fixed bosonic mass fraction, this can lead either to extended, 
halo-like configurations around a small fermionic core, or to more compact 
mixed configurations in which both components occupy comparable spatial 
regions. The former exist at low total masses but fall below the 
observational range highlighted in the figure and are not shown in our 
plots; they should not be mistaken for forbidden solutions, since mixed 
configurations with $M_b/M_T \lesssim 10\%$ do exist at low total mass. 
Only at high total masses do variations of the bosonic mass fraction 
produce a measurable spread in $R_e$, so that the width of the two-dimensional 
solution band expands and the colored region broadens near the 
maximum-mass points. This organization is qualitatively consistent with 
static fermion-boson stars in GR \cite{DiGiovanni:2021ejn} and with the 
static $f(R)$ study of \cite{2018PhRvD..97b4030L}. 

Before comparing our equilibrium solutions with observations, we
briefly recall the physical content of each constraint shown in
Fig.~\ref{fig9}.  The NICER measurements of PSR~J0030$+$0451
(PSRA, $M\approx1.34\,M_{\odot}$, $R\approx12.71\,\mathrm{km}$) \cite{Riley:2019yda}
and PSR~J0740$+$6620 (PSRB, $M\approx2.08\,M_{\odot}$,
$R\approx12.4\,\mathrm{km}$) \cite{Legred:2021hdx} anchor the mass-radius relation at
low-to-intermediate and high masses respectively, providing direct
simultaneous constraints on stellar radius and mass from X-ray
pulse-profile modeling.  The gravitational-wave event GW170817
(GWB) \cite{LIGOScientific:2017vwq} constrains the tidal deformability of merging NSs
and, through post-merger dynamics, supports the upper-mass bound
of Rezzolla et al.\ (RMW) \cite{Rezzolla:2017aly} and
the tension analysis of Nathanail et al.\ (NMR) \cite{Nathanail:2021tay}, which together
delimit the expected maximum compact-star mass for nucleonic models.
Also, GW190425 (GWC, $M_{\mathrm{tot}}\sim3.40\,M_{\odot}$) \cite{LIGOScientific:2020aai} extends this
picture to higher masses, while GW190814 (GWA,
$M_{\mathrm{secondary}}\sim2.50$-$2.67\,M_{\odot}$) \cite{LIGOScientific:2020zkf} is the most
discriminating constraint: its secondary component falls within the
black-hole low-mass gap, and whether it represents an exceptionally
massive compact star or a light black hole has profound implications
for both nuclear physics and gravitational theory.

The static and Keplerian sequences shown in each panel of
Fig.~\ref{fig9} bracket the space of uniformly rotating equilibrium
configurations with $0\leq\Omega\leq\Omega_{\rm K}$. Most of the
overplotted observational bounds, notably the NICER ellipses and the
GWB tidal-deformability constraints, refer to nonrotating or
slowly rotating compact stars. The comparison with these data should
therefore focus on configurations close to the static limit, rather
than on the rapidly rotating portion of the Keplerian branch. Within
this interpretation, all three gravitational theories exhibit regions of solution space that intersect the relevant constraints. Because
observed pulsar spins remain well below mass shedding
($\lesssim700\,\mathrm{Hz}$ for the fastest known systems), the
associated rotational corrections to the mass-radius relation are
expected to be modest in the astrophysically relevant spin range.

The PSRA and PSRB ellipses are intersected by equilibrium configurations 
in all three models for small bosonic mass fractions, indicating that in 
this regime the AkmalPR EOS largely governs the global structure. Since 
the observational constraints lie between the static and mass-shedding 
limits, a continuous family of uniformly rotating configurations is 
expected to be compatible with these measurements, with the static and 
Keplerian sequences providing natural bounds. The NICER sources 
PSRA (PSR~J0030+0451, $\nu \approx 205\,\mathrm{Hz}$) and PSRB 
(PSR~J0740+6620, $\nu \approx 346\,\mathrm{Hz}$) rotate at frequencies 
that are significant but still well below the Keplerian limit for typical 
NS models; the associated rotational corrections to the mass-radius 
relation are therefore modest, and the relevant equilibrium configurations 
are expected to lie close to, but not exactly at, the static limit. The 
bosonic field enlarges the accessible parameter space, allowing additional 
equilibrium configurations that still intersect the NICER ellipses at low $M_b/M_T$. 

Moreover, the three models show good consistency with the RMW and NMR observational constraints over a range of rotation rates, from the static limit up to the Keplerian sequence. Notably, the $R$-squared gravity models cover a broader range of the allowed parameter space compared to the corresponding GR solutions. Similarly, the GWC band can be justified by moderately rotating configurations in the three
cases at radii $R_{e}\approx11$-$16\,\mathrm{km}$.
 
The most physically significant differentiation among the three gravitational
scenarios emerges precisely at the GWA constraint. We note that the 
difficulty of accounting for a secondary mass in the range suggested by 
GW190814 is not specific to the AkmalPR EOS adopted here; it is a robust 
feature of GR-based models with conventional nuclear matter, as discussed 
in \cite{Nathanail:2021tay, Rezzolla:2017aly}. In the GR case,
the static maximum mass falls \emph{below}
the lower edge of the GWA band, so that a non-rotating or very slowly 
rotating fermion-boson star in GR cannot account for the
GW190814 secondary mass; reaching this mass range in GR requires higher 
angular velocity solutions, which simultaneously inflates the equatorial 
radius to $R_{e}\gtrsim14\,\mathrm{km}$, creating a residual tension with
the moderate radii preferred by PSRB. The inclusion of either 
$R$-squared gravity or a bosonic component, or both, provides a natural 
mechanism to alleviate this tension without requiring extreme assumptions 
about the nuclear interaction. 

The $a=10$ model partially
resolves this tension: its static maximum of $2.51\,M_{\odot}$
just reaches the lower edge of the GWA band, and its
maximum mass spans
a wide range of GWA at moderate spin rates well below
$\Omega_{\mathrm{K}}$, reducing the radius tension. The
advantage is most pronounced for $a=10^{4}$: its static maximum
mass lies
\emph{within} the GWA band, making it the most favourable scenario in which
a genuinely non-rotating mixed fermion-boson star with a sub-$10\%$
bosonic fraction can reproduce the GWA secondary mass without
invoking any particular spin. 

\section{Conclusions}\label{sec:vi}

We have presented a new family of equilibrium solutions describing mixed fermion-boson star
configurations within the framework of $R$-squared gravity, $f(R) = R + a R^2$, for both static and rotating regimes. Besides obtaining fermion-boson star solutions in GR, we have constructed
the corresponding models for two representative $R$-squared models governed by the parameter $a$, including a
regime in which the theory effectively approaches a Brans-Dicke-like behavior. The fermionic sector is described
by the realistic AkmalPR EOS, while the bosonic component is treated within the RPV
approximation, which allows one to recast the bosonic energy-momentum tensor in a perfect-fluid form amenable to
the self-consistent field method. The resulting numerical scheme, implemented as an extension of the \texttt{RNS} code, preserves the general philosophy of stationary axisymmetric equilibrium
construction through elliptic equations while incorporating the scalar degree of freedom as an additional effective source. This can be considered as a self-consistent framework for constructing rotating mixed compact stars in a viable $f(R)$ theory of gravity. 
 
Our results show that $R$-squared gravity, treated through an equivalent scalar-tensor representation with a dynamical scalar degree of freedom, exerts a systematic influence on the internal structure of both the fermionic and bosonic sectors. For fixed global quantities, namely identical total mass, angular momentum, and internal bosonic frequency, the inclusion of the scalar degree of freedom leads to a suppression of the bosonic field amplitude and a redistribution of the fermionic energy density toward more spatially extended profiles. Moreover, the results indicate that this behavior reflects a modification of the effective gravitational potential experienced
by the bosonic matter, which reduces the localized field concentration required to support a configuration of
fixed total mass. The scalar-induced corrections are most pronounced in the dense central regions of the star,
where curvature is strongest and the coupling between the scalar degree of freedom and the matter fields is most
effective, and they weaken progressively toward the outer stellar boundary.
 
Our analysis of rotating configurations reveals that the bosonic sector develops a toroidal topology
characteristic of spinning boson stars, with the bosonic field vanishing along the rotation axis and reaching
its maximum at the equatorial plane. The dimensionless internal frequency $\bar{w}$ acts as the primary
control parameter governing whether the bosonic component remains confined within the fermionic interior as
a core-like structure or extends outward to form a halo surrounding the NS component. We have found
that increasing parameter $a$ systematically enlarges the region of the parameter space accessible to compact or even core-like
mixed configurations, enabling equilibrium solutions with different bosonic fractions at lower internal frequencies
that are simply not present in GR. 

The mass-radius sequences constructed for static configurations and for sequences at the Keplerian mass-shedding limit show that the $R$-squared term increases both limiting maximum masses relative to GR, with a consistently larger relative increase at the Keplerian boundary than at the static one. This statement refers only to the sequence maxima in Fig.~\ref{fig9} and should not be read as a general claim about rotating models: in the fixed-$(M_T,J_T,\bar w)$ profile analysis of the previous section, a larger angular momentum partially suppresses the modified-gravity-induced changes in the central fermionic energy density.

We have confronted our equilibrium sequences with current astrophysical and gravitational-wave constraints, including NICER measurements of PSR~J0030+0451 and PSR~J0740+6620, the tidal-deformability bounds from GW170817, and the secondary-component mass inferred from GW190814. Our findings show that the three studied gravitational scenarios remain broadly compatible with these observational constraints, while the $R$-squared gravity models display a richer and broader spectrum of solutions than the corresponding GR case. The most physically discriminating difference emerges in connection with the GW190814 constraint: within the AkmalPR EOS adopted here, no nonrotating or slowly rotating mixed star in GR can account for the secondary mass without invoking near-Keplerian rotation. We note that this is not a general theoretical bound, as sufficiently stiff nuclear EOSs can in principle reach $2.6\,M_\odot$ in GR \cite{Nathanail:2021tay, Rezzolla:2017aly}; however, simultaneously satisfying the PSRB radius constraint makes this combination difficult to achieve. By contrast, the $R$-squared gravity models can accommodate this mass even within near-static sequences. These results suggest that mixed fermion-boson stars in $R$-squared gravity may provide a viable stellar interpretation for compact objects lying within the black-hole low-mass gap, while remaining within a framework that can be directly confronted with gravitational-wave observations. 

Although the present analysis provides a robust and self-consistent framework for constructing rotating fermion-boson stars in $R$-squared gravity, several natural extensions could further broaden the scope of this work. The RPV approximation adopted for the bosonic sector is well motivated and has been validated in related studies. Nevertheless, extending the analysis beyond the high-coupling limit would allow one to quantify the contribution of the exponentially decaying scalar-field tail, particularly in halo-dominated configurations where its effects may become non-negligible. In addition, while the equilibrium solutions obtained here are physically consistent, a dedicated perturbative stability analysis would provide important complementary information regarding the dynamically stable branches of the solution space, especially in the high-mass regime where both the scalar degree of freedom and bosonic sectors play significant roles. The assumption of uniform rotation adopted throughout this work is standard and appropriate for the present study; however, extending the framework to differentially rotating configurations could reveal additional families of equilibrium solutions and enrich the associated phenomenology. On the matter side, a systematic investigation employing different nuclear equations of state would help disentangle modified-gravity effects from uncertainties associated with nuclear physics.
 
Finally, investigating the role of dynamical gravitational synchronization between the fermionic and bosonic sectors, as recently explored in \cite{Lazarte:2025etl},  also represents a well-motivated extension of the current framework, to study how such a mechanism modifies the equilibrium sequences constructed here and their oscillation frequency spectra. Pursuing these directions may further strengthen the astrophysical relevance of mixed fermion-boson stars in modified gravity as viable compact-object candidates and as promising sources for current and future gravitational-wave observations. 

\section{Acknowledgements}
The authors thank Alejandro Cruz Osorio for his contribution regarding the treatment of observational bounds. SF is funded by the Conselleria de Innovación, Universidades, Ciencia y Sociedad Digital of the Generalitat Valenciana and the European Social Fund through a postdoctoral fellowship APOSTD 2025 (CIAPOS/2024/461). Partial support of KP-06-N62/6 from the Bulgarian science fund is also gratefully acknowledged. DD acknowledges financial support via an Emmy Noether Research Group funded by the German Research Foundation (DFG) under grant no. DO 1771/1-1, by the Spanish Ministry of Science and Innovation via the Ram\'on y Cajal programme (grant RYC2023-042559-I), funded by MCIN/AEI/10.13039/501100011033 and by ESF+. NSG acknowledges support from the Spanish Ministry of Science and Innovation via the Ram\'on y Cajal programme (grant RYC2022-037424-I), funded by MCIN/AEI/10.13039/501100011033 and by ``ESF Investing in your future”. JCM is funded by a UNAM-DGAPA Postdoctoral Fellowship. This work is also supported by the Spanish Agencia Estatal de Investigación (grant PID2024-159689NB-C21) funded by MICIU/AEI/10.13039/501100011033 and by FEDER/EU, by the Generalitat Valenciana (Prometeo grant CIPROM/2022/49), and by the European Horizon Europe staff exchange (SE) programme HORIZON-MSCA2021-SE-01 Grant No. NewFunFiCO-101086251.

\bigskip

\bibliography{draft_ml} 

\begin{appendix}
\section{Static Spacetime}
\label{app:static_spacetime}
This appendix collects the static and spherically symmetric form of the fermion-boson system in $R$-squared gravity. This formulation is useful for comparison with standard static fermion-boson-star models and for clarifying the nonrotating limit of the configurations discussed in the main text. We emphasize, however, that the equations below are written in Schwarzschild-like coordinates, whereas the rotating configurations constructed in this paper use quasi-isotropic coordinates. Therefore, the static metric presented here should not be interpreted as a direct coordinate-by-coordinate limit of the rotating line element in Eq.~\eqref{metric1}.

The static and spherically symmetric metriz ansatz is written as
\begin{equation}\label{metric}
ds^2 = -N(r)^2 dt^2 +L(r)^2 dr^2 + r^2(d\theta^2 + \sin^2 \theta d\psi^2).
\end{equation}
The energy-momentum tensor of the fermionic part is given by Eq.~(\ref{SE_f}) and its 4-velocity is set by choosing $\Omega=0$. Correspondingly, the bosonic field ansatz, reads
\begin{equation}
\Phi_{\rm s}(t,r)=\phi_{\rm s}e^{-i w_{\rm s}t}
\end{equation}
where the label ``s" indicates that this ansatz corresponds to the static configuration. This field is composed by a radial part $\phi_{\rm s}=\phi_{\rm s}(r)$ and by the harmonic time dependence, containing one of the key parameters of the problem, the field frequency $w_{\rm s}$. In the standard static treatment, one fixes the central amplitude $\phi_{\rm s}(0)=\phi_0$, imposes regularity of the metric and matter fields at the center and asymptotic flatness at infinity, and determines the field frequency $w_{\rm s}$ as an eigenvalue of the resulting boundary-value problem. In the present work we do not follow this procedure. Within the RPV approximation, the bosonic field is fixed algebraically by Eq.~\eqref{ryan_field}, and the static configurations (shown in \Cref{sec:v}) are obtained as the $J_T \to 0.0 M_{\odot}^2$ limit of the rotating numerical scheme, as described in \Cref{section_III_b}. 

On the other hand, the additional source terms arising from the $R$-squared modification can be moved to the right-hand-side of Eq.~\eqref{Einstein:tens}. These contributions may then be interpreted as an effective stress-energy tensor associated with the scalar degree of freedom formulation.

Under such assumptions, the equations that describe static fermion-boson stars in $R$-squared gravity take the following form:
\begin{widetext}
\begin{align}
\frac{dL}{dr} &= \frac{L}{2}\biggl\{\left( \frac{1 -L^2}{r}\right)
+ 8\pi r\left[A^4(\varphi) L^2(\rho_b + \rho_f)\right]
+ r\left(\frac{d\varphi}{dr}\right)^2
+ \frac{L^2r}{2}V(\varphi)\biggr\}, \label{dadr} \\[6pt]
\frac{dN}{dr} &= \frac{N}{2}\biggl\{\frac{L^2 - 1}{r}
+ 8\pi r\left[A^4(\varphi) L^2 (p_{b} + p_{f})\right]
+ r\left(\frac{d\varphi}{dr}\right)^2
- \frac{L^2r}{2}V(\varphi)\biggr\}, \label{dbdr} \\[6pt]
\frac{d\Psi}{dr} &= -\left[\frac{1}{N}\frac{dN}{dr}-\frac{1}{L}\frac{dL}{dr}+\frac{2}{r}+2\kappa(\varphi)\frac{d\varphi}{dr}\right]\Psi
+ L^2\left[A^2(\varphi)(\mu_b^2 + \lambda\phi_{s}^2)
- \frac{w_s^2}{N^2}\right]\phi_{s}, \\[6pt]
\frac{d\vartheta}{dr} &= -\left(\frac{1}{N}\frac{dN}{dr}-\frac{1}{L}\frac{dL}{dr}+\frac{2}{r}\right)\vartheta + L^2\bigg\{\frac{1}{4}\frac{dV}{d\varphi}
+ 4\pi \kappa(\varphi) \left[A^{4}(\varphi)\left(\rho_f - 3p_f + 2\mu_b^2\phi_s^2 +\lambda \phi_{s}^4\right) - A^2(\varphi)\left(\frac{w_s^2\phi_s^2}{N^2}\right) -\frac{\Psi^2}{L^2}\right]\bigg\}, \\[6pt]
\frac{dp_{f}}{dr}
&= -(\rho_{f}+p_{f})\left(\frac{1}{N} \frac{dN}{dr} + \kappa(\varphi) \frac{d \varphi}{dr}\right),
\end{align}
\end{widetext}
where $\Psi=d\phi_{s}/dr$ and $\vartheta=d\varphi/dr$. Although the Klein-Gordon equation is intrinsically a second-order differential equation, we recast it into an equivalent first-order system by introducing an auxiliary radial variable. This formulation is standard in the literature on boson stars and mixed fermion-boson stars, as it facilitates numerical implementation while maintaining consistency with the conventional notation.

\section{Field equations for rotating fermion-boson stars}
\label{appx_source}

This appendix reports the explicit form of the field equations we solve to build the rotating fermion-boson star models discussed in this work (including the source terms $S_{T}^{x}(r,\mu)$). These equations are implemented and solved in our modified version of the \texttt{RNS} code, discussed in \Cref{sec:iii}. In the Komatsu-Eriguchi-Hachisu scheme, the elliptic equations for the metric potentials $\gamma$, $\rho$, $\sigma$, and $\alpha$, together with the equation for the gravitational scalar degree of freedom $\varphi$, are written with a linear differential operator on the left-hand side and effective sources on the right-hand side. Besides the fermionic and bosonic matter contributions, these sources contain the $R$-squared terms associated with $\varphi$ and $V(\varphi)$. The explicit expressions, including the radial and angular derivatives of the metric functions and matter variables, are given below
\begin{widetext}
\begin{align}
\left(\Delta + \frac{1}{r}\partial_r - \frac{\mu}{r^2}\partial_\mu\right)\left(\gamma e^{\gamma/2}\right)
&= e^{\gamma/2}
\Bigg\{
\left[16\pi A^4(\varphi)(p_f + p_b) - V(\varphi)\right]e^{2\alpha}
\nonumber\\
&\quad
+\frac{\gamma}{2}
\Big[
\left[16\pi A^4(\varphi) (p_f + p_b) - V(\varphi)\right]e^{2\alpha}
-\frac12(\partial_r\gamma)^2
-\frac12\frac{1-\mu^2}{r^2}(\partial_\mu\gamma)^2
\Big]
\Bigg\},
\label{eq:DiffEq_gamma}
\end{align}

\begin{align}
\Delta(\rho e^{\gamma/2})
&= e^{\gamma/2}
\Bigg\{
8\pi e^{2\alpha} A^4(\varphi)\left[(\rho_{f}+p_{f})\frac{1+\upsilon^2}{1-\upsilon^2} + (\rho_{b}+p_{b})\frac{1+\bar{\upsilon}^2}{1-\bar{\upsilon}^2}\right]
+r^2(1-\mu^2)e^{-2\rho}
\Big[(\partial_r\sigma)^2
+\frac{1-\mu^2}{r^2}(\partial_\mu\sigma)^2\Big]
\nonumber\\
&\quad
+\frac{1}{r}\partial_r\gamma
-\frac{\mu}{r^2}\partial_\mu\gamma
+\frac{\rho}{2}
\Big[
\left[16\pi A^4(\varphi) (p_{f} + p_{b})-V(\varphi)\right]e^{2\alpha}
-\frac{1}{r}\partial_r\gamma
+\frac{\mu}{r^2}\partial_\mu\gamma -\frac12(\partial_r\gamma)^2
-\frac12\frac{1-\mu^2}{r^2}(\partial_\mu\gamma)^2
\Big]
\Bigg\},
\label{eq:DiffEq_sigma}
\end{align}

\begin{align}
\left(\Delta+\frac{2}{r}\partial_r-\frac{2\mu}{r^2}\partial_\mu\right)
\left(\sigma e^{\gamma/2-\rho}\right)
&= e^{\gamma/2-\rho}
\Bigg\{
-16\pi e^{2\alpha} A^4(\varphi) \left[\frac{(\rho_f+p_f)(\Omega-\sigma)}{1-\upsilon^2}+\frac{\Bar{\upsilon}\left(\rho_{b}+p_{b}\right)}{(1-\Bar{\upsilon}^2)r\sin\theta}\right]
\nonumber\\
&
+\sigma\Big[
-\frac{1}{r}\partial_r\!\left(\frac{\gamma}{2}+2\rho\right)
+\frac{\mu}{r^2}\partial_\mu\!\left(\frac{\gamma}{2}+2\rho\right)
-\frac14(\partial_r\gamma)^2
-\frac14\frac{1-\mu^2}{r^2}(\partial_\mu\gamma)^2
\nonumber\\
&
+(\partial_r\rho)^2
+\frac{1-\mu^2}{r^2}(\partial_\mu\rho)^2
-r^2(1-\mu^2)e^{-2\rho}
\Big[(\partial_r\sigma)^2
+\frac{1-\mu^2}{r^2}(\partial_\mu\sigma)^2\Big]
\nonumber\\
&
-8\pi e^{2\alpha} A^4(\varphi) \left[\frac{(1+\upsilon^2)\rho_f+2\upsilon^2 p_{f}}{1-\upsilon^2}+\frac{(1+\Bar{\upsilon}^2)\rho_{b}+2\Bar{\upsilon}^2 p_{b}}{1-\Bar{\upsilon}^2}\right]
-\frac{1}{2}V(\varphi)e^{2\alpha}
\Big]
\Bigg\},
\label{eq:DiffEq_omega}
\end{align}

\begin{align}
\Delta \varphi=&-\partial_{r}\gamma\partial_{r}\varphi -\frac{1-\mu^2}{r^2}\partial_{\mu}\gamma\partial_{\mu}\varphi +\Big\{-\frac{4\pi}{\sqrt{3}}A^{4}(\varphi)\left[\rho_{b}+\rho_{f} - 3(p_{b}+p_{f})\right]+\frac{1}{4}\frac{dV(\varphi)}{d\varphi}\Big\}e^{2\alpha},
\end{align}

\end{widetext}
The operator $\Delta$ is defined by
\begin{equation}
\Delta \equiv \partial_r^2 + \frac{2}{r}\partial_r + \frac{1-\mu^2}{r^2}\partial_\mu^2 - \frac{2\mu}{r^2}\partial_\mu .
\end{equation}

For the metric potential $\alpha$, two coupled first–order partial differential equations can be derived. Nevertheless, within our numerical implementation, only one of these relations is required to determine $\alpha$, which we take it to be

\begin{widetext}
\begin{align}
\partial_\mu \alpha
&=
-\frac{1}{2}\left(\partial_\mu \gamma+\partial_\mu \rho\right)
-
\Big\{(1-\mu^2)(1+r\partial_r\gamma)^2
+\big[-\mu+(1-\mu^2)\partial_\mu\gamma\big]^2\Big\}^{-1}
\nonumber\\
&\times
\Bigg\{
\frac12
\Big[
r\partial_r(r\partial_r\gamma)
+r^2(\partial_r\gamma)^2
-(1-\mu^2)(\partial_\mu\gamma)^2
-\partial_\mu\!\left[(1-\mu^2)\partial_\mu\gamma\right]
+\mu\partial_\mu\gamma
\Big]
\big[-\mu+(1-\mu^2)\partial_\mu\gamma\big]
\nonumber\\
&\qquad
+\frac14
\big[-\mu+(1-\mu^2)\partial_\mu\gamma\big]
\Big[
r^2(\partial_r\gamma+\partial_r\rho)^2
-(1-\mu^2)(\partial_\mu\gamma+\partial_\mu\rho)^2
+4r^2(\partial_r\varphi)^2
-4(1-\mu^2)(\partial_\mu\varphi)^2
\Big]
\nonumber\\
&\qquad
+\mu r\partial_r\gamma(1+r\partial_r\gamma)
-(1-\mu^2)r(1+r\partial_r\gamma)
\Big[
\partial_\mu\partial_r\gamma
+\partial_\mu\gamma\,\partial_r\gamma
+\frac12(\partial_\mu\gamma+\partial_\mu\rho)
(\partial_r\gamma+\partial_r\rho)
+2\partial_\mu\varphi\,\partial_r\varphi
\Big]
\nonumber\\
&\qquad
+\frac14(1-\mu^2)e^{-2\rho}
\Big[
-\big[-\mu+(1-\mu^2)\partial_\mu\gamma\big]
\big(r^4(\partial_r\sigma)^2-r^2(1-\mu^2)(\partial_\mu\sigma)^2\big)
+2(1-\mu^2)r^3
\partial_\mu\sigma\,\partial_r\sigma
(1+r\partial_r\gamma)
\Big]
\Bigg\}.
\label{eq:DiffEq_alpha}
\end{align}
\end{widetext}
\section{Set 2 solutions}
\label{appendix0}
\begin{figure*}[!ht]
 \centering
 \includegraphics[angle=0,width=1\hsize]{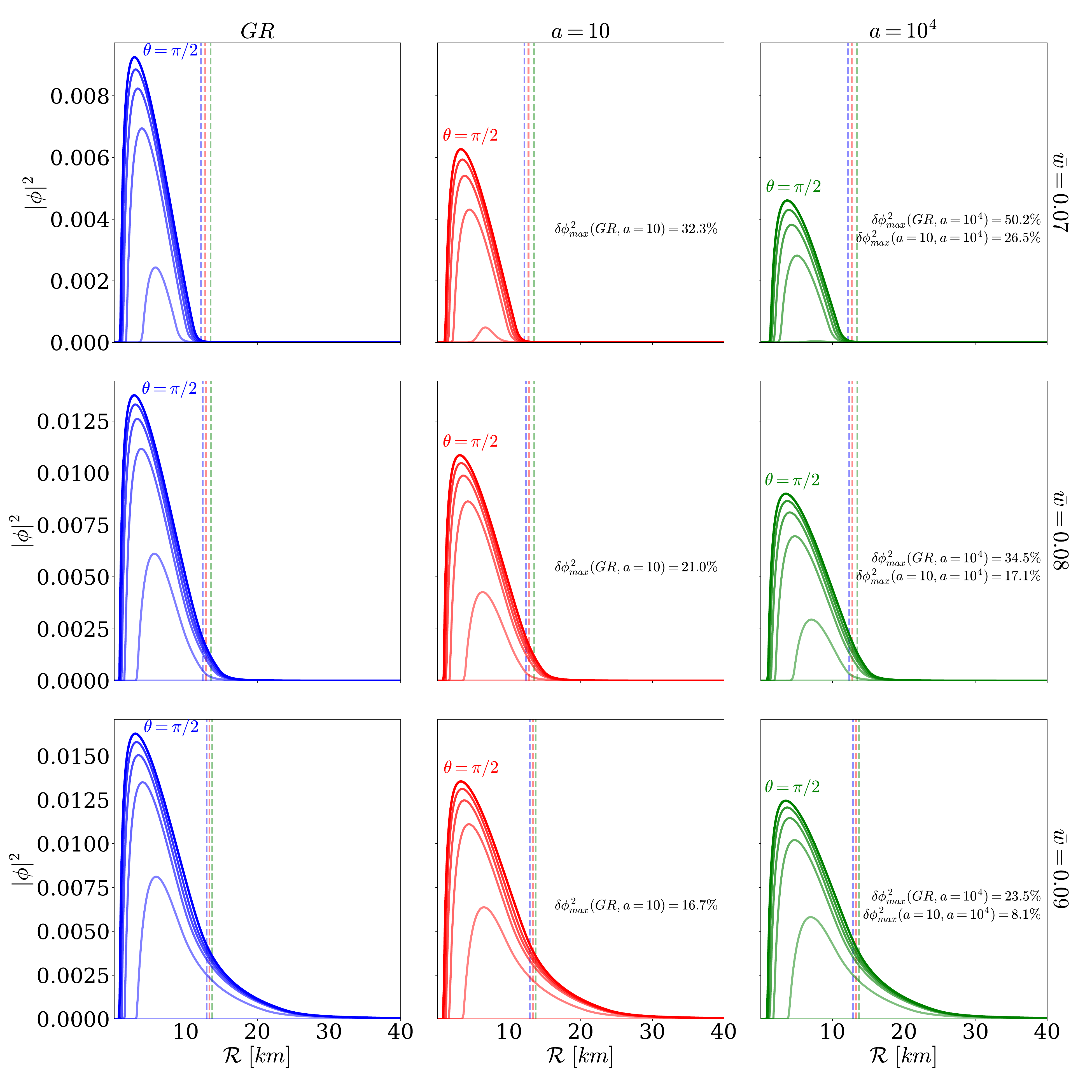}
 \caption{Similar to Fig.\,\ref{fig2}, but for $J_{T}=2.0$ and $M_{T}=2.0 M_{\odot}$.}
\label{fig4}
\end{figure*}

\begin{figure*}[!ht]
 \centering
  \includegraphics[angle=0,width=1\hsize]{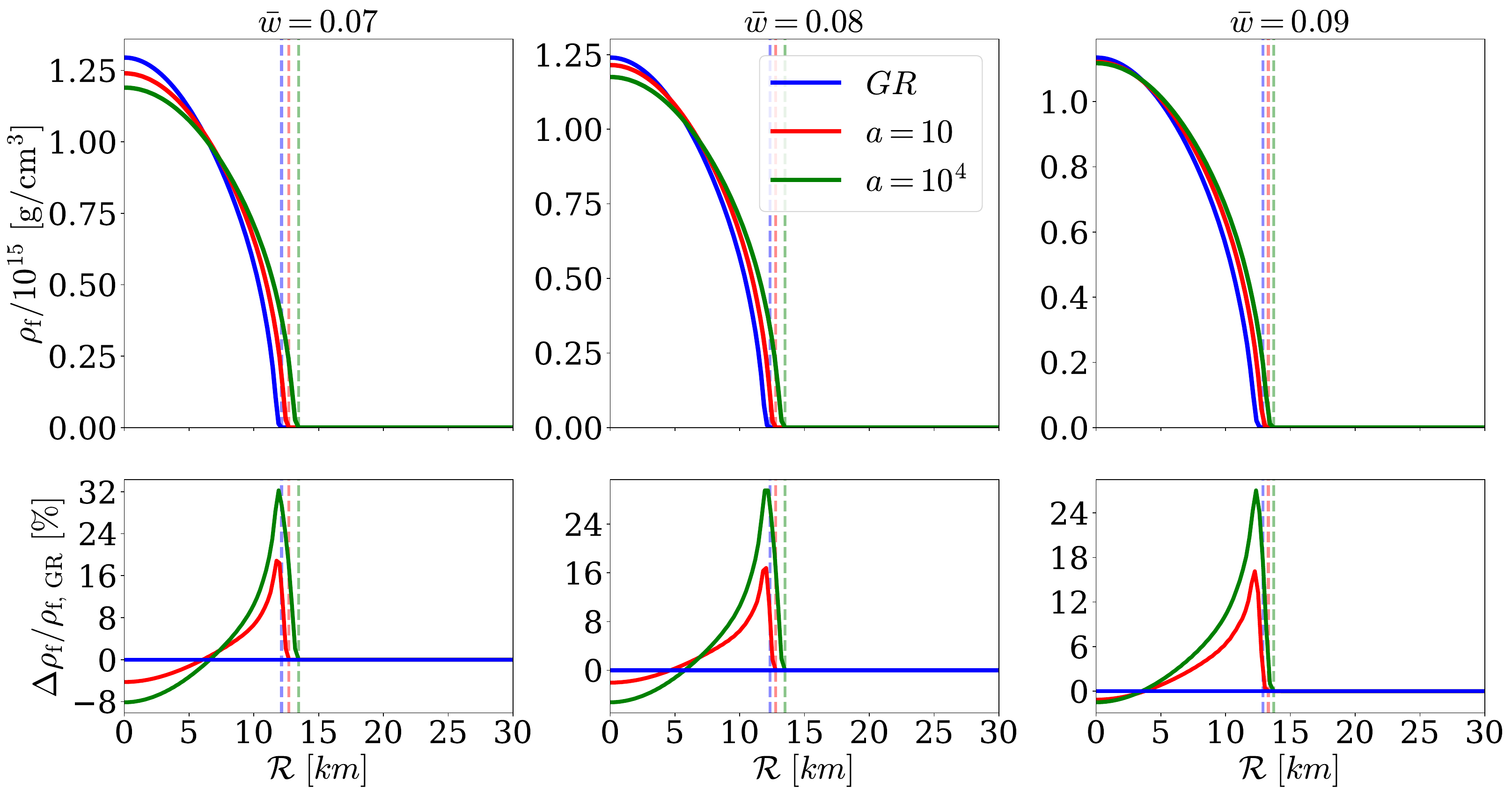}
 \caption{Similar to Fig.\,\ref{fig3}, but for $J_{T}=2.0 M_{\odot}^2$ and $M_{T}=2.0 M_{\odot}$.}
\label{fig5}
\end{figure*}

Figs.~\ref{fig4} and~\ref{fig5} show the results obtained for the Set 2 of fermion-boson stars, displaying the radial profile of the bosonic field and the fermionic energy density, respectively, for the different gravitational theories and fixed values of the internal bosonic frequency. As discussed in the main text, the qualitative trends are the
same as those found for Set~1.
 
We recall that the distinction between the two sets lies in the value of the
fixed angular momentum, which is twice as large for Set 2.
This increase in angular momentum has a visible impact on the structure of
the solutions. In particular, the angular dependence of both the bosonic field
and fermionic energy-density profiles becomes more pronounced as the angular momentum
increases, reflecting the stronger departure from spherical symmetry in the
more rapidly rotating configurations.

The bosonic field profiles shown in Fig.~\ref{fig4} exhibit the same frequency dependent behavior observed in Set~1. For $\bar{w}=0.07$, the bosonic component is strongly localized within the fermionic core, forming a compact distribution. As $\bar{w}$ increases, the bosonic field profile becomes progressively broader, eventually developing a halo-like structure that extends well beyond the fermionic surface at $\bar{w}=0.09$. The scalar degree of freedom continues to suppress the peak amplitude of
$|\phi|^2$ relative to GR across all frequency values, with the $a = 10^4$
model yielding the largest deviation, as annotated in each panel.

Compared to Set~1, however, the maximum relative deviations between the GR and $R$-squared gravity models are slightly reduced for all three values of $\bar{w}$. This should not be read as a contradiction with the previous statement that rotation can amplify departures from GR. Here $M_T$, $J_T$, and $\bar{w}$ are held fixed while the gravitational theory is varied; in that comparison, a larger angular momentum partially suppresses the profile-level
departures from GR through the additional centrifugal support. By contrast, in
the mass-radius analysis of \Cref{sec:v} one compares the static and
Keplerian boundaries of the equilibrium sequences, and there rotation
systematically raises the maximum supported mass, so that the relative
enhancement due to $R$-squared gravity is larger at the Keplerian limit than
at the static one. The present result therefore reflects a competition between
centrifugal and scalar-field effects at fixed global parameters, not a general
statement that rotation always weakens modified-gravity signatures. This behavior indicates that the higher angular momentum of Set~2 introduces a
stronger centrifugal contribution to the equilibrium, which partially
counteracts the scalar-induced modifications to the bosonic sector. Nevertheless, the overall qualitative behavior remains unchanged. The presence of the scalar field systematically reduces the degree of field localization required to sustain a configuration with fixed total mass. This influence gradually weakens as $\bar{w}$ increases and the bosonic component evolves from a compact core-like configuration toward a more extended halo-like structure.

On the other hand, the fermionic energy-density profiles in Fig.~\ref{fig5}
exhibit a behavior fully consistent with that of Set 1. In both modified-gravity
scenarios, the central energy density $\bar{\rho}_f(0)$ decreases
systematically relative to the GR baseline, accompanied by a localized
positive enhancement at intermediate radii within the fermionic core. This
pattern reflects a redistribution of baryonic matter over a moderately
enlarged stellar volume, driven by the modified balance between gravity and
pressure support mediated by the scalar degree of freedom. The percentage deviations shown
in the bottom panels confirm that the $a = 10^4$ model produces the largest
departures from GR, while the $a = 10$ case yields intermediate corrections.
A notable quantitative difference with respect to {Set~1} is the overall
suppression of the fermionic energy-density deviations in { Set~2}. The larger
angular momentum strengthens the centrifugal support of the stellar
structure, reducing the sensitivity of the fermionic distribution to the scalar field. For the specific combination of mass and angular momentum studied here, this behavior suggests that the rotational contribution to the equilibrium configuration may partially counteract the scalar-induced corrections, though a more systematic survey of the parameter space would be needed to establish this as a general trend. As a result, the fermionic sector appears less sensitive to variations in the modified-gravity parameter $a$ than in the slower-rotation regime of {Set~1}. As in the static and { Set~1} sequences, the scalar-field effects on $\bar{\rho}_f$ are strongest at lower values of $\bar{w}$, where the fermionic and bosonic components are spatially co-located. These effects gradually weaken as the bosonic component develops a more extended halo surrounding the compact fermionic core. 

\section{Supplementary analyses}
\label{appendixb}

\begin{figure*}[!ht]
 \centering
 \includegraphics[angle=0,width=1\hsize]{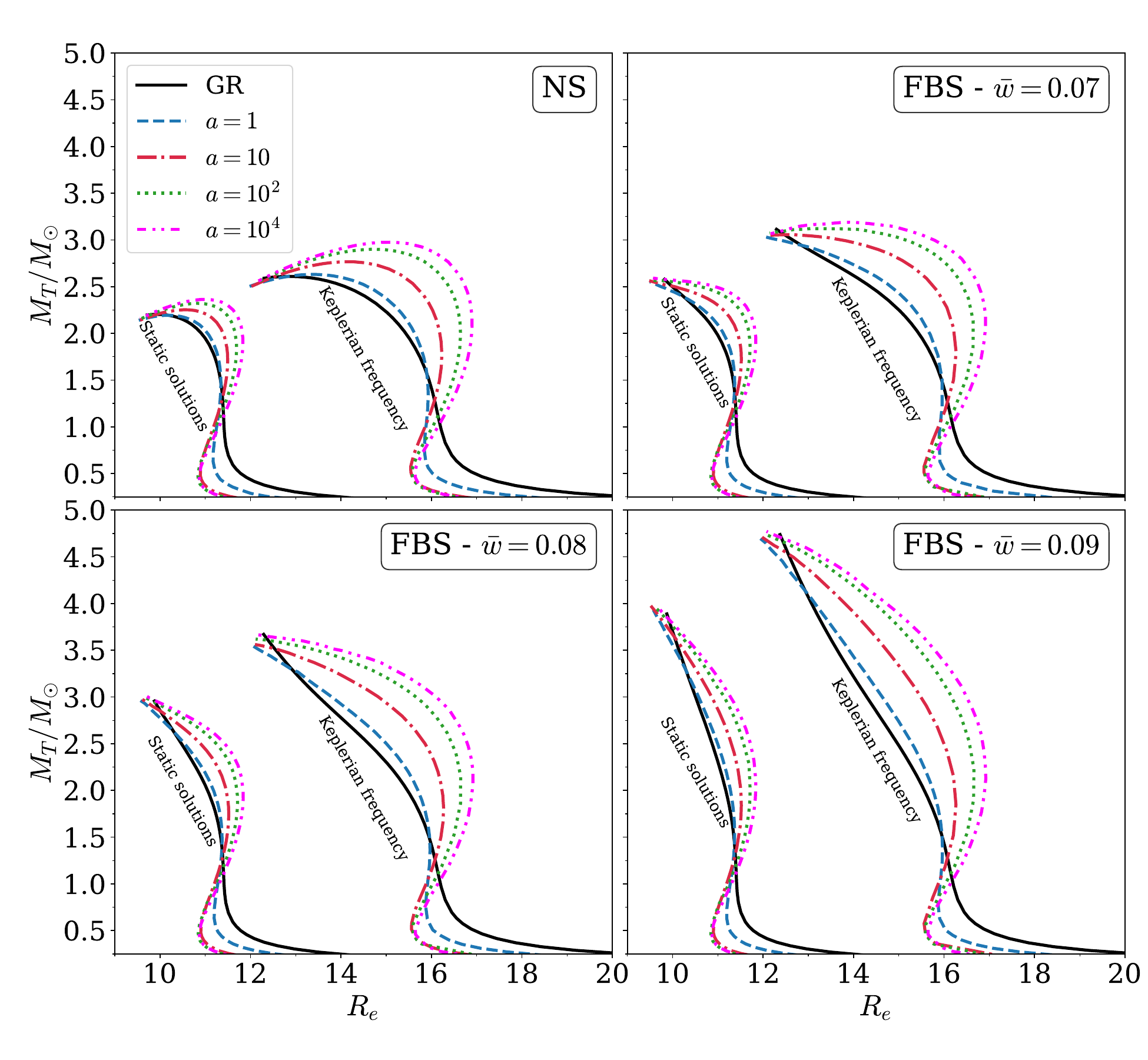}
 \caption{The mass-radius relation for AkmalPR EOS in the case of static stars and stars rotating at the Kepler limit. Different styles and colors of the curves correspond to different values of the parameter $a$. }
\label{fig6}
\end{figure*}

For completeness, we present here supplementary analyses that complement the discussion of the results in \Cref{sec:v} of the main text.

First, Fig.~\ref{fig6} displays mass-radius diagrams for five gravitational theories, including GR and $R$-squared models with $a=1, 10, 10^2$, and $10^4$. Unlike the analysis in the main text, the internal bosonic frequency is now fixed in each panel, and the resulting equilibrium sequences are compared across gravity theories. Both static and Keplerian limits are shown for all cases. The four panels correspond to a pure NS configuration and mixed fermion-boson stars with $\bar{w}=0.07$, $0.08$, and $0.09$. Smaller $\bar{w}$ produces more compact bosonic profiles concentrated within the fermionic core, while larger values lead to more extended, halo-like configurations.

Increasing the parameter $a$ may lead to an expansion of both the static and Keplerian mass-radius regions for all configurations, which can indicate the robustness of the scalar-induced mass enhancement. At the same time, the separation between curves decreases for larger $a$, indicating a saturation of modified-gravity effects. This saturation arises because large $a$ corresponds to a very small scalar field mass so that the theory approaches a Brans-Dicke-like limit in which the scalar field becomes effectively massless and its influence on the stellar structure reaches a maximum that cannot be further increased by raising $a$. As noted in the main text, the value $a=10^4$ was chosen precisely to probe this near-saturation regime, representing close to the maximum deviation from GR accessible within $R$-squared gravity \cite{Yazadjiev:2015zia, Doneva:2015hsa}. The inclusion of the bosonic field further enlarges the accessible mass-radius space, particularly in the Keplerian regime, where it contributes significantly to the total mass. For fixed $\bar{w}$, the solution family is mainly governed by the central fermionic energy density, with the bosonic amplitude adjusting accordingly as $\rho_c$ increases. 

\begin{figure*}[!ht]
 \centering
 \includegraphics[angle=0,width=1\hsize]{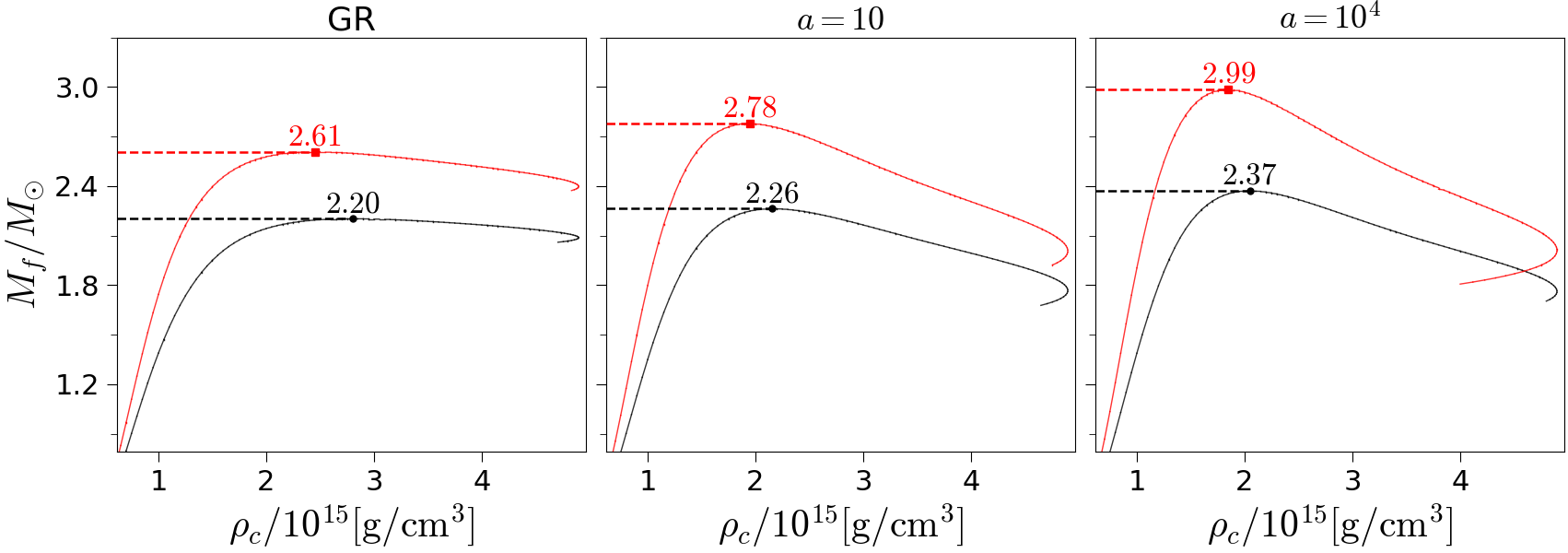}
\caption{Fermionic core mass of the mixed star configurations, $M_f$, as a function of the central fermionic energy density, $\rho_c$, within the context of GR and two $R$-squared gravity models with $a=10$ and $a=10^4$. Black curves correspond to static equilibrium sequences and red curves to the Keplerian mass-shedding limit. Filled markers indicate the maximum-mass configurations, with numerical values annotated in each panel.} 
\label{fig8}
\end{figure*}

Despite the fixed-frequency constraint, these solutions occupy an extreme region of the mass-radius diagram where both the scalar gravitational degree of freedom and the bosonic component play a dominant role. The resulting mixed configurations consist of a massive fermionic core embedded within an extended bosonic halo. Although a dedicated stability analysis is still required, these solutions can potentially represent transitional stages in gravitational collapse or may act as possible progenitors of scalarized black holes. Any connection to scalarized black-hole solutions should, however, be treated with caution. In scalarized black holes, the existence of scalar hair requires synchronization between the scalar-field configuration and the black-hole angular velocity \cite{Herdeiro:2014goa,Herdeiro:2015gia, Doneva:2022ewd}, whereas no such synchronization condition is imposed on the mixed-star equilibria considered here, where the fermionic and bosonic sectors remain dynamically independent.

\begin{figure*}[!ht]
 \centering
 \includegraphics[angle=0,width=1\hsize]{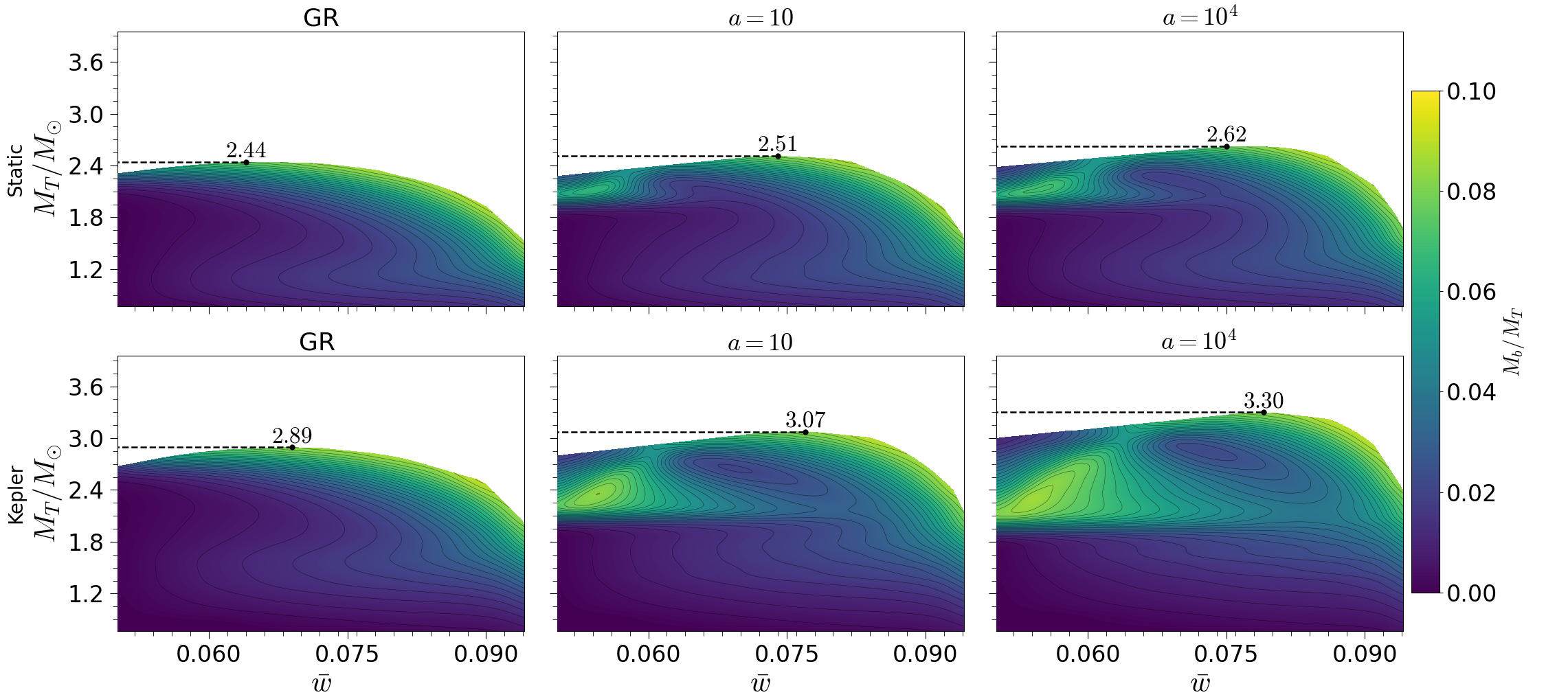}
 \caption{Total mass as a function of the bosonic field frequency within the framework of GR and two distinct $R$-squared gravity models, shown for static equilibrium ({\bf top}) and the Keplerian limit ({\bf bottom}). The different colors in the density regions represent the fractional contribution of the bosonic component to the total gravitational mass. }
\label{fig_fin}
\end{figure*}

Fig.~\ref{fig8} shows the dependence of the fermionic gravitational mass $M_f$ on the central fermionic energy density $\rho_c$ for  three gravitational theories. In all cases, both the static and Keplerian sequences exhibit the standard behavior characteristic of relativistic stellar configurations: $M_f$ increases with $\rho_c$ up to a well-defined maximum and then decreases, indicating that configurations beyond the turning point are typically associated with dynamical instability. The inclusion of the scalar degree of freedom systematically enhances the maximum fermionic mass and shifts the location of the peak toward lower central densities as the parameter $a$ increases. Quantitatively, the maximum static fermionic mass rises from $2.20\,M_\odot$ in GR to $2.26\,M_\odot$ for $a = 10$ and $2.37\,M_\odot$ for $a = 10^4$, while the corresponding Keplerian values increase from $2.60\,M_\odot$ in GR to $2.77\,M_\odot$ and $2.98\,M_\odot$, respectively. Moreover, for larger values of $a$, the peak of the $M_f$-$\rho_c$ relation becomes noticeably sharper, suggesting an increased sensitivity of the fermionic mass to variations in the central energy density within the $R$-squared models.

Finally, Fig.~\ref{fig_fin} shows a series of diagrams illustrating the dependence of the total mass $M_T$ on the bosonic internal frequency $\bar{w}$. The panels are arranged such that the upper row corresponds to static configurations, while the lower row shows the corresponding Keplerian-limit solutions. The three columns follow the same ordering adopted in the main analysis, namely GR, $a=10$, and $a=10^4$. The contour region represents the bosonic mass fraction $M_b/M_T$, whereas contour lines delineate iso-fraction curves of constant bosonic-to-total mass ratio. In each sequence, the maximum attainable total mass is indicated by a horizontal dashed line. Across all three gravitational frameworks, the results reveal a characteristic dome-shaped structure in the mass-frequency plane, where the total mass increases with $\bar{w}$ up to a peak and subsequently decreases at higher frequencies. 

A clear restructuring of the solution space emerges when comparing GR with the $R$-squared gravity cases. In GR, the mass distribution remains relatively smooth, with larger bosonic fractions primarily confined to the high-mass regime. By contrast, once the scalar degree of freedom is introduced, the morphology of the solution space changes substantially. In particular, the scalar field induces a systematic upward shift of the entire mass-frequency surface, and the maximum total mass increases monotonically with the parameter $a$ in both static and Keplerian configurations. More significantly, the $R$-squared models exhibit localized high-intensity regions in the color map at comparatively lower values of $\bar{w}$ and $M_T$, indicating the emergence of compact configurations with enhanced bosonic fractions that are absent in GR.

This feature becomes particularly noticeable in the mass range ($2.0 \lesssim M_T/M_\odot \lesssim 3.0$), especially at lower bosonic frequencies. Interestingly, this interval overlaps with observationally suggested mass windows, which may point to some astrophysical relevance. At the same time, it lies within a region where unstable branches of mixed configurations are also expected to appear. Overall, the solution space becomes more diverse as $a$ increases, indicating that the scalar field might play an important role in shaping both the global structure and the range of fermion-boson star configurations.

\end{appendix}

\end{document}